\documentclass{article}
\usepackage[utf8]{inputenc}

\usepackage{chngcntr}
\usepackage{graphicx}   %
\usepackage{float}      %
\usepackage{caption}    %
\usepackage{subcaption} 
\usepackage{booktabs}       %
\usepackage{chemformula} %
\usepackage{amsfonts}       %
\usepackage[version=4]{mhchem}
\usepackage[numbers,sort&compress]{natbib} %
\usepackage{authblk}    %

\usepackage{lineno} %

\usepackage[normalem]{ulem}

\usepackage[parfill]{parskip} %

\usepackage{hyperref}
\usepackage{xcolor}
\hypersetup{
  colorlinks   = true, %
  urlcolor     = red, %
  linkcolor    = blue, %
  citecolor   = blue %
}

\begin{document}

\title{Higher-Order Equivariant Neural Networks
for Charge Density Prediction in Materials}

\author[1*]{Teddy Koker}
\author[1]{Keegan Quigley}
\author[1,2]{Eric Taw}
\author[1]{Kevin Tibbetts}
\author[1*]{Lin Li}
\affil[1]{MIT Lincoln Laboratory, Lexington, MA 02421}
\affil[2]{Material Science Division, Lawrence Berkeley National Laboratory, Berkeley, CA 94720}
\affil[*]{Correspondence to \texttt{\{thomas.koker, lin.li\}@ll.mit.edu}}
\date{}

\date{}

\newcommand{\MODEL}{ChargE3Net}

\maketitle

\begin{abstract}

The calculation of electron density distribution using density functional theory (DFT) in materials and molecules is central to the study of their quantum and macro-scale properties, yet accurate and efficient calculation remains a long-standing challenge. We introduce \MODEL{}, an E(3)-equivariant graph neural network for predicting electron density in atomic systems. \MODEL{} enables the learning of higher-order equivariant feature to achieve high predictive accuracy and model expressivity. We show that \MODEL{} exceeds the performance of prior work on diverse sets of molecules and materials. When trained on the massive dataset of over 100K materials in the Materials Project database, our model is able to capture the complexity and variability in the data, leading to a significant 26.7\% reduction in self-consistent iterations when used to initialize DFT calculations on unseen materials. Furthermore, we show that non-self-consistent DFT calculations using our predicted charge densities yield near-DFT performance on electronic and thermodynamic property prediction at a fraction of the computational cost. Further analysis attributes the greater predictive accuracy to improved modeling of systems with high angular variations. These results illuminate a pathway towards a machine learning-accelerated ab initio calculations for materials discovery.

\end{abstract}

\section{Introduction}

Quantum mechanical interactions of electrons in atoms provide a foundation for understanding and predicting properties of materials~\cite{PhysRev.136.B864}. Modern density functional theory (DFT), particularly the Kohn-Sham (KS) formulation of DFT~\cite{PhysRev.140.A1133}, is by far the most widely used method for performing such electronic structure calculations due to its balanced tradeoff between accuracy and computational cost. Because KS-DFT obtains the electronic structure by iteratively solving Kohn-Sham equations and re-calculating the electron charge density until convergence to a self-consistent field (SCF), extensive use of DFT is hindered by the unfavorable $O(N^3)$ scaling with the system size, making it infeasible for large-scale calculations beyond a few hundred atoms. 

While prior works on linear-scaling DFT~\cite{prenticeONETEPLinearscalingDensity2020,solerSIESTAMethodInitio2002,mohrAccurateEfficientLinear2015}, local coupled cluster methods~\cite{riplingerEfficientLinearScaling2013,riplingerNaturalTripleExcitations2013}, and orbital-free formulation of DFT~\cite{wittOrbitalfreeDensityFunctional2018} have nominally reduced the scaling limit, large-scale calculation is still unobtainable for general problems. In recent years, many efforts have focused on the potential of machine learning (ML) approaches to overcome the quantum scaling limit of DFT. ML models have been developed to map chemical and material systems to properties to enable high-throughput screening~\cite{xie2018crystal, chen2019graph, louis2020graph, dunnBenchmarkingMaterialsProperty2020,choudhary2021atomistic,koker2022graph}. Once trained, ML models are computationally fast, but they are restricted to properties that already exist in a database (typically at the ground state), limiting their applicability to general problems. ML models have also been proven useful for predicting atomic energies and forces for large-scale molecular dynamic simulations~\cite{drautzAtomicClusterExpansion2019,batzner20223,musaelianLearningLocalEquivariant2023, chmielaMachineLearningAccurate2017,du2023machine}. The process generally requires a dedicated computational campaign with DFT to generate high quality data for a subset of materials with a pre-defined potential or temperature range \cite{chmielaMachineLearningAccurate2017,chanussotOpenCatalyst20202021,tranOpenCatalyst20222023,sriramOpenDAC20232023,pengmeiMD17ReactiveXxMD2024,dengCHGNetPretrainedUniversal2023,du2023machine}. A more recent work has trained a foundation model across the periodic table for predicting interatomic potentials \cite{batatiaFoundationModelAtomistic2024}, but it still requires additional DFT data and model finetuning for applications beyond structure relaxation. An alternative approach is to directly predict converged DFT quantities (e.g., electron charge density) to reduce the number of SCF iterations needed to solve KS equations. Since the number of SCF iterations is dependent on the accuracy of the initialized charge density, a fast and accurate charge density prediction may be able to bypass the need for self-consistent iterations, enabling accurate computation of downstream properties at a fraction of the computational cost.

Several works have used ML for charge density prediction. One approach is to expand the charge density with a local atomic basis set and use ML to predict the expansion coefficients~\cite{brockherde2017bypassing,grisafi2018transferable,fabrizio2019electron,lewis2021learning,qiao2022informing, rackers2023recipe, del2023deep}. 
Local atomic orbital basis sets are often parametrized for specific elements. While these models are in principle more efficient, scaling on the order of the number of basis functions, they are restricted to the basis set they are trained on, thus impacting generalizability. Furthermore, any model that attempts to find the parameters of atomic basis sets that best fit grid-based data is fundamentally limited by the expressivity of the basis set~\cite{del2023deep}. A parallel line of research focuses on the prediction of the Hamiltonian operator~\citep{liDeeplearningDensityFunctional2022, unke2021se, nigam2022equivariant, zhong2024universal}, however these are also restricted by the choice of basis set, and often not obtainable through DFT codes. 

An alternative approach is to learn the electron charge density directly from a discretized grid of density points in 3D Cartesian space. The grid-based approach is agnostic to the basis set (plane-wave or atomic basis) used for the quantum chemistry calculation and are a natural input format for DFT codes~\citep{kresse1993initio,kresse1996efficiency,kresse1996efficient,giannozzi2009quantum}. Existing grid-based methods~\citep{gong2019predicting,jorgensen2020deepdft,jorgensen2022equivariant,sunshineChemicalPropertiesGraph2023,focassio2023linear} have shown promising results on small molecules and small subsets of material classes. Pioneering works~\cite{gong2019predicting, jorgensen2020deepdft} have focused on learning the charge density using graph neural networks invariant to translation, rotation and permutation~\cite{xie2018crystal, schutt2017schnet}. \citet{jorgensen2020deepdft} later demonstrate an improvement in accuracy using an equivariant graph neural network, PaiNN~\citep{jorgensen2022equivariant}, through $\mathbb{R}^3$ vector representations and rotationally equivariant operations \citep{schutt2021equivariant}. While these models have been shown to achieve high accuracy on small, specialized molecular or material datasets, no work has demonstrated applicability on larger, diverse datasets across the periodic table. 
In addition, these models use scalar and/or vector representations to construct rotationally equivariant networks, which could be further improved by incorporating higher-order equivariant features, as has been shown in other tasks, such as the prediction of atomic forces~\citep{batzner20223}. Finally, limited work has demonstrated the utility of ML-predicted charge densities to initialize DFT calculations, with existing work showing only marginal reduction of SCF steps \citep{pope2024towards,sunshineChemicalPropertiesGraph2023}.

In this work, we develop a grid-based model, \MODEL{}, that uniquely combines graph neural networks, SO(3) equivariant features, and higher-order tensor operations, to accurately predict the electron charge density. Our model only uses atomic species and positions as input without the need for additional features. It enables the use of higher-order equivariant features to achieve more accurate performance and higher model expressivity. In a head-to-head comparison with prior grid-based and basis function-based methods, we show that our method not only achieves greater accuracy on existing benchmark data, but also is expressive enough to capture the complexity and variability present in Materials Project (MP) data \cite{shen2022representation}, consisting of more than 100K inorganic materials across the periodic table and for all materials classes. We demonstrate the application of MP-trained \MODEL{} to initialize DFT calculations, resulting in a median of 26.7\% reduction in SCF steps on unseen MP data and 28.6\% on novel GNoME materials \cite{merchant2023scaling}. Both have surpassed the reductions demonstrated by prior work \citep{pope2024towards,sunshineChemicalPropertiesGraph2023}. When evaluating the utility of \MODEL{} in computing a wide range of electronic and thermodynamic properties non-self-consistently, we show a great portion of materials achieving near-DFT accuracy. Further analysis on materials with less-accurate predictions provides insights into a potential cause of the observed inconsistency between self-consistent and non-self-consistent calculations to inform future research. By enabling higher-order equivariant models, we can also investigate the impact of equivariant representations in model performance.
We show that higher-order representations yield more accurate charge density predictions for materials with high angular variation. Lastly, We demonstrate the linear time complexity of our model with respect to 
system size, showing the capability of predicting density on systems with $>10^4$
atoms, surpassing what is feasible with ab initio calculations.

\begin{figure}
    \centering
    \includegraphics[width=\linewidth]{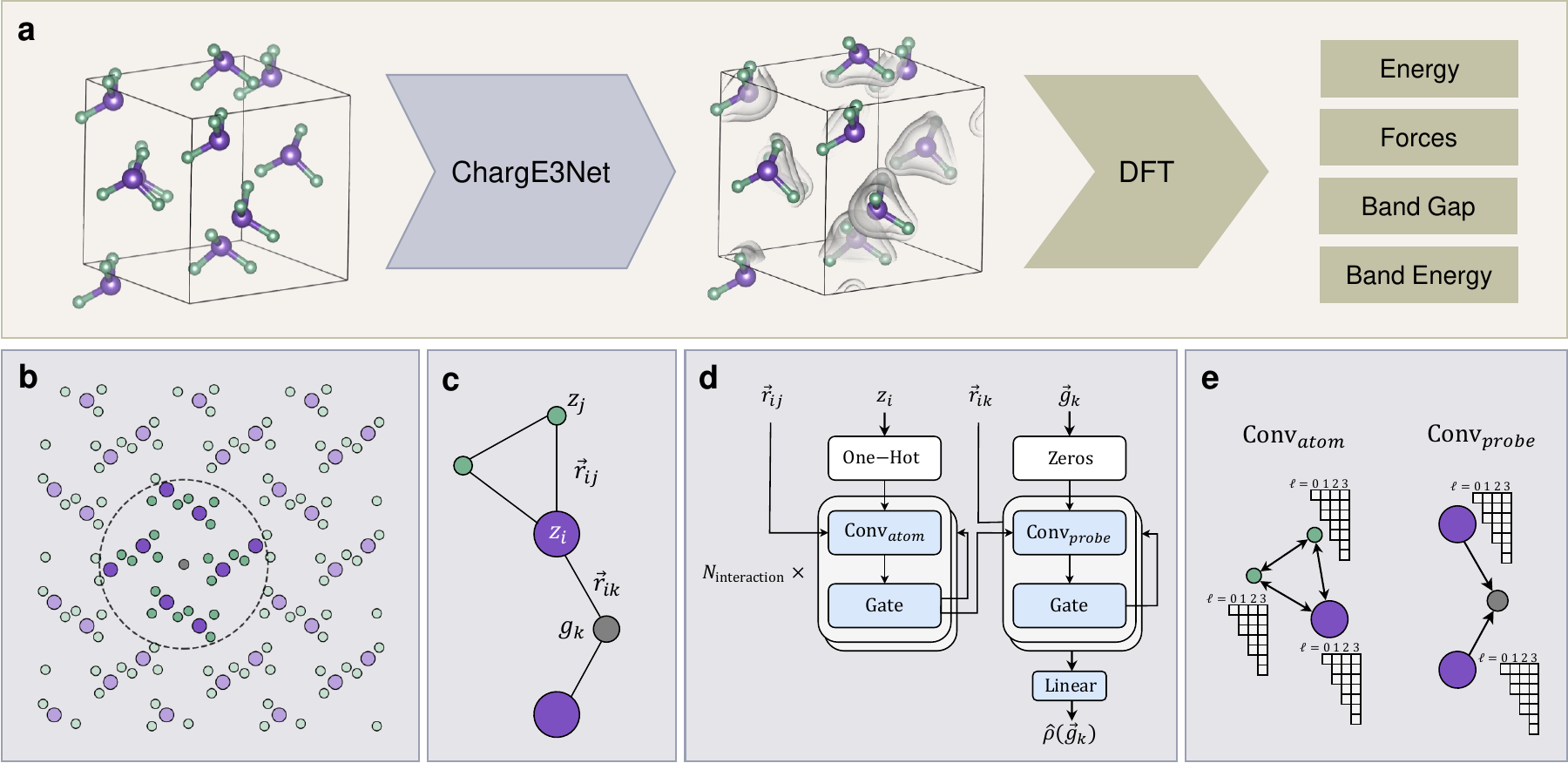}
    \caption{
        \textbf{Overall architecture of the \MODEL{} model.}
        \textbf{a} Acceleration of DFT property prediction with ChargE3Net charge density initialization. \textbf{b} Illustration of single charge density probe point (center, gray) and its local environment within a periodic atomic system. \textbf{c} Sub-graph of local environment graph with atom nodes $z_i$, $z_j$ and probe node $g_k$. \textbf{d} Neural network architecture. Predicted charge density $\hat{\rho}(\vec{g})$ at probe point $\vec{g}_k$ is computed from the atomic species $z_i$, interatomic  displacement vectors $\vec{r}_{ij}$, and probe-atom displacement vectors $\vec{r}_{ik}$ through a series of convolution and gate operations. \textbf{e} $\text{Conv}_\text{atom}$ sends higher-order representations bidirectionally to atom nodes. $\text{Conv}_\text{probe}$ sends higher-order representations from atoms to probes.
      }
    \label{fig:architecture}
\end{figure}

\section{Results}

\subsection{Development of higher-order equivariant neural networks for charge density prediction} \label{equivariance}

The charge density, along
with other properties such as forces, is equivariant under rotation and
translation of the atomic system. That is, a rotation or translation of
the atomic system in Euclidean space will result in an equivalent rotation or
translation to the charge density or forces. Formally, this is known as
equivariance with respect to E(3), which includes rotations, translations, and
reflections in 3D space. This can be achieved in principle by using only
invariant scalar features \cite{schutt2017schnet},
 such as inter-atomic
distances; however, this prevents the model from using angular information, limiting the
accuracy that can be attained~\cite{pozdnyakov2022incompleteness}.
More recent work \cite{schutt2021equivariant} has introduced equivariance methods using vector $\mathbb{R}^3$ features, such as relative atomic
positions, to incorporate angular information, and has shown to improve the performance of charge density prediction \cite{jorgensen2022equivariant}. We hypothesize that higher-order representations could help improve model performance and accelerate DFT calculations.

ChargE3Net achieves equivariance through means outlined by tensor field
networks~\cite{thomas2018tensor}. 
Translation equivariance is achieved by using
relative atomic positions; rotation equivariance is achieved by restricting
features to be irreducible representations, \textit{irreps}, of SO(3), which are operated upon by equivariant functions. These
features take the form $V^{(\ell, p)}_{cm}$, a dictionary of tensors with keys representing rotation
order $\ell \in \{0, 1, 2, ...\}$
and parity $p \in \{-1, 1\}$. Each tensor
has a channel index $c \in [0, N_\text{channels})$ and an index $m \in [-\ell, \ell]$.
In this way, the
representation at a given rotation order $\ell$ and parity $p$ would have a size of $\mathbb{R}^{N_\text{channels}\times (2\ell+1)}$.
These representations are combined with the equivariant tensor product operation $\otimes$,
using Clebsch–Gordan coefficients $C$~\cite{thomas2018tensor} as
implemented in \texttt{e3nn} \cite{e3nn_paper}: %

\begin{equation} \label{tp}
    \left(U^{(\ell_1,p_1)} \otimes V^{(\ell_2,p_2)}\right)^{(\ell_o,p_o)}_{cm_o} =
    \sum_{m_1=-\ell_1}^{\ell_1}
    \sum_{m_2=-\ell_2}^{\ell_2}
    C^{(\ell_o, m_o)}_{(\ell_1,m_1)(\ell_2,m_2)}
    U_{cm_1}^{(\ell_1,p_1)}
    V_{cm_2}^{(\ell_1,p_2)}
\end{equation}

where $\ell_o$ and $p_o$ are given by $|\ell_1 - \ell_2| \leq \ell_o \leq | \ell_1 + \ell_2 |$ and 
$p_o=p_1p_2$.
We maintain only those representations with $\ell_o \leq L$
where $L$ is a maximum allowed rotation order. 
For ease of notation, we will omit the keys $\ell$, $p$ and indices $c$, $m$ for these tensor representations for the rest of the paper. Unless otherwise indicated, we use a $L=4$ maximum rotation order for \MODEL{}.

Fig.~\ref{fig:architecture} shows the model architecture of ChargE3Net. To predict charge density of a given atomic system, a set of $M$ probe points $\{\vec{g}_1, ..., \vec{g}_M\} \in \mathbb{R}^3$ are introduced to locations where charge densities are to be predicted. We represent predicted
charge density $\hat{\rho}(\vec{g})$ as an equivariant graph neural network with inputs of atomic
numbers $\{z_1, ..., z_N\} \in \mathbb{N}$ with respective locations of the atoms
$\{\vec{r}_1, ..., \vec{r}_N\} \in \mathbb{R}^3$ where $N$ denotes the total number of atoms, as well as probe locations
$\{\vec{g}_1, ..., \vec{g}_M\} \in \mathbb{R}^3$ and an optional periodic boundary cell $B\in\mathbb{R}^{3 \times 3}$ for periodic systems. 

As illustrated in Fig.~\ref{fig:architecture}b and \ref{fig:architecture}c, a graph is first constructed with atoms and
probe points as vertices, with edges formed via proximity with a cutoff of 4~\AA. Edges connecting atoms are unidirectional and edges connecting atoms and probe points are directed pointing from atoms to probe points. Vertex features for atoms $\{A_i\}_{i=0}^{N}$ are initialized as a one-hot encoding of the atomic number, represented as a $\ell=0$, $p=1$ tensor with $N_{\rm channels}$ equal to the number of atomic species. Features for probe points $\{P_k\}_{k=0}^M$ are initialized as a single scalar zero (i.e., $\ell=0$, $p=1$ and $N_{\rm channels}=1$).

ChargE3Net uses a graph neural network (GNN) to update these tensor features (Fig.~\ref{fig:architecture}d). Within each layer of GNN, the network performs message passing between neighboring atoms and probes, where probe points can only receive messages from neighboring atoms (Fig.~\ref{fig:architecture}e). The core of the message passing in ChargE3Net is a convolution function that updates the vertex tensor feature with neighboring vertices by the convolution operation (Eqn.~\eqref{eqn:convolution} in Methods). Since the proposed convolution operation replaces the linear message passing with equivariant operation through the tensor product defined in Eqn.~\eqref{tp}, the operation is rotation and translation equivariant; see Methods for more detail. 

ChargE3Net conducts a series of equivariant convolution and gate operations to obtain representations for each atom, with an addition series of equivariant convolution and gate operations to obtain representations for each probe point.
The learned tensor feature associated to each probe point is passed to a post-processing layer where a simple regression is used to predict the charge density at the corresponding probe location $\hat{\rho}(\vec{g}_k)$.

\subsection{Performance evaluation}
We train and validate ChargE3Net on a diverse set of DFT-computed charge density data, including 
organic molecules (QM9)
\cite{ruddigkeit2012enumeration,ramakrishnan2014quantum,qm9vasp}, nickel manganese cobalt battery cathode materials (NMC)
\cite{nmcvasp}, and inorganic materials collected from Materials Project (MP)
\cite{jain2013commentary,shen2022representation}. The first two datasets have been used in the past for benchmarking charge density models for small molecules and a subset of crystal systems with boundary conditions~\citep{jorgensen2022equivariant,qiao2022informing}. To develop a charge density model for the entire periodic table, we investigate one of the largest open databases of DFT, the Materials Project (MP) database. It includes charge densities for over 100K bulk materials with diverse compositions and structures (Supplementary Fig.~\ref{fig:tsne}-\ref{fig:radarplot} and Methods). 

We assess the performance our models using mean absolute error normalized by the total number of electrons within a volume of the atomic system~\cite{grisafi2018transferable,fabrizio2019electron,jorgensen2022equivariant,qiao2022informing},  $\epsilon_\text{mae}$, which is computed via numerical integration on the charge density grid points $G$:

\begin{equation}\label{eq:mae}
\epsilon_\text{mae} = \frac
{\sum_{\vec{g_k} \in G}\left|\rho (\vec{g_k})-\hat{\rho }(\vec{g_k})\right|}
{\sum_{\vec{g_k} \in G}\left|\rho (\vec{g_k})\right|}
\end{equation}

Unless otherwise noted, the probe points $G$ hereby reference the full set of discretized unitcell grid points for which we have DFT-computed charge density values. 

\begin{table}[!t]
  \caption{Model performance for charge density prediction}
  \label{table:nmape}
  \centering
  \begin{tabular}{lllll}
    \toprule
    Dataset           & invDeepDFT      & equiDeepDFT              & OrbNet-Equi              & \MODEL{} \\
    \midrule
    NMC               & $0.089\pm0.001$ & $\mathbf{0.061}\pm0.001$ & -                        & $\mathbf{0.060}\pm0.001$\\
    QM9               & $0.357\pm0.001$ & $0.284\pm0.001$          & $0.206\pm0.001$          & $\mathbf{0.196}\pm0.001$\\
    MP                & $0.859\pm0.011$ & $0.799\pm0.010$          & -                        & $\mathbf{0.523}\pm0.010$ \\
    \bottomrule
  \end{tabular}
  {\raggedright \small{ All values are presented as an average $\epsilon_\text{mae}$ (\%) and $\pm$ one standard error. Bold values represent the best $\epsilon_\text{mae}$ (\%) scores. $\epsilon_\text{mae}$ reported for MP include both magnetic and non-magnetic materials.} \par}
\end{table}

Table~\ref{table:nmape} compares the performance of \MODEL{} to models introduced in OrbNet-Equi~\citep{qiao2022informing} and DeepDFT~\citep{jorgensen2022equivariant}. For QM9 and NMC datasets, we use identical training, validation, and test splits as \citet{jorgensen2022equivariant}. All reported $\epsilon_\text{mae}$ are computed using publicly released models and verified to match the numbers reported in the original work. For the MP dataset, we train the DeepDFT models using the same experimental settings from the original work (see Supplementary section for more detail). We observe that all equivariant models perform better than invariant models, demonstrating the importance of respecting rotation equivariance. Moreover, ChargE3Net model significantly outperforms prior equivariant DeepDFT models \citep{jorgensen2022equivariant} on the Materials Project (MP)
and QM9 datasets, while achieving similar performance on the NMC dataset. This suggests that the inclusion of higher-order representations can improve model performance. More remarkably,
our model achieves a lower $\epsilon_\text{mae}$ on the QM9 dataset than OrbNet-Equi \citep{qiao2022informing} model, which leverages additional features based on quantum mechanical calculations while \MODEL{} only uses atomic species and position 
information.
This shows how the model is capable of learning highly informative representations of quantum interactions in a purely data-driven approach.

\MODEL{}'s performance on the MP data demonstrates that it is capable of successfully modelling charge density across the periodic table and a wide variety of material classes that are not present in other datasets. When we compare the prediction performance with respect to the number of unique species in a material, the MP-trained \MODEL{} is capable of predicting charge density in systems with many unique species, without a significant change in error distribution as number of species increases (Supplementary Fig.~\ref{fig:nmape_species}).
This suggests \MODEL{} can reproduce a broad range of chemical interactions at the electronic level without incurring the cost of an ab initio calculation.

\subsection{Using predicted charge densities to accelerate DFT}

DFT calculations require an initial charge density and wavefunction prior to iteratively solving the Kohn-Sham equations until self-consistency \cite{sholl2009density}. 
Charge density initializations that closely match the ground-state charge density often save compute time by reducing the number of self-consistent field (SCF) steps needed to find a self-consistent solution. 
Many schemes for initializing the charge density exist in the literature, but the most common one is the superposition of atomic densities (SAD), where the initial charge density is computed by summing over the atomic charge density of each atom in the system~\cite{lehtola2019assessment}. 

\begin{figure}[t!]
    \centering
    \includegraphics[width=\linewidth]{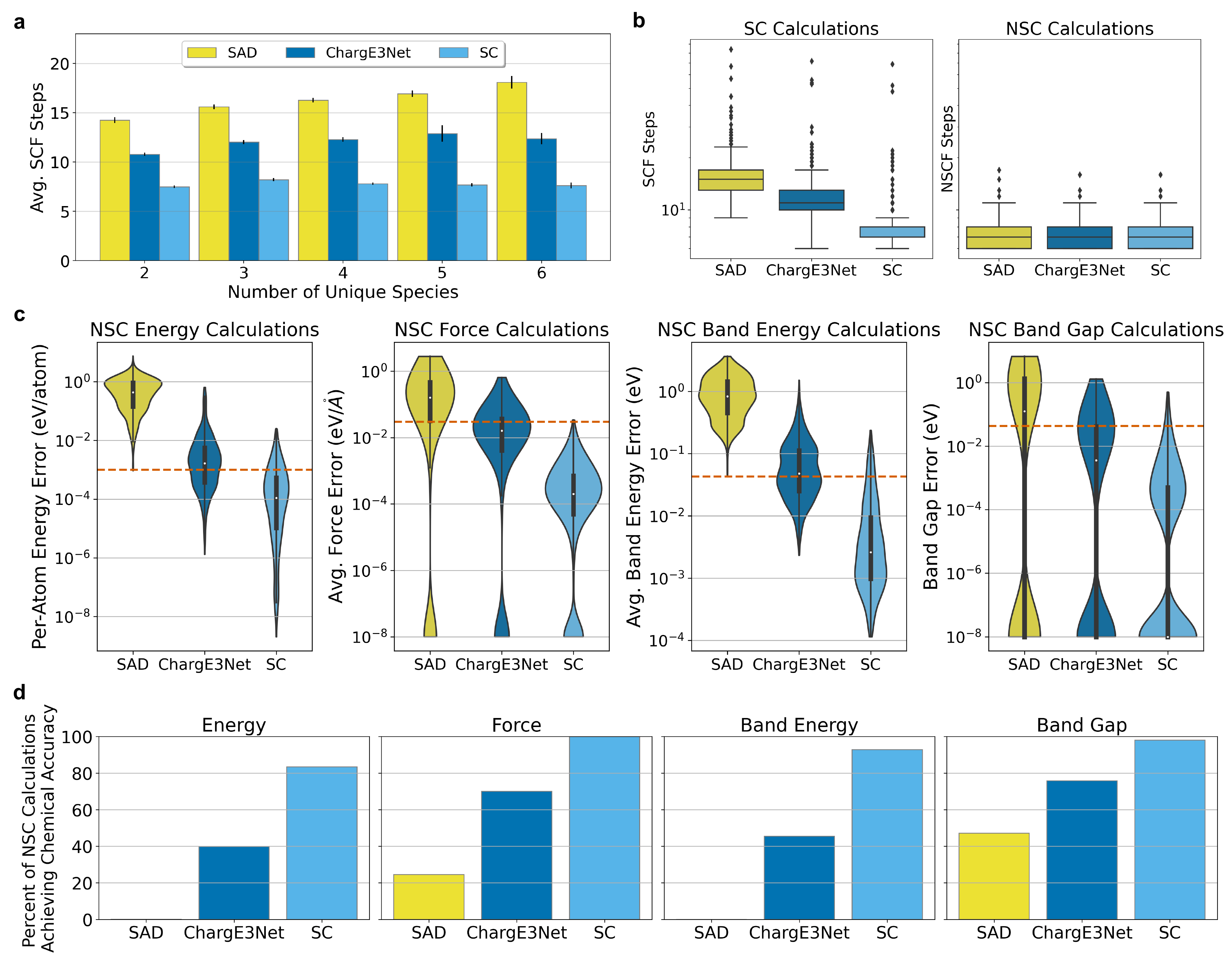}
    \caption{\textbf{DFT calculations with \MODEL{}.} Comparison of DFT calculations initialized from superposition of atomic densities (SAD), \MODEL{} predictions, and ground-truth densities from existing self-consistent calculations (SC).
    \textbf{a} Average number of SCF steps required to converge the total energy to $5\times 10^{-5}$ eV/atom for non-magnetic materials in the MP test set, with respect to the number of unique species within each material. Error bar shows $\pm$ one standard error.
    \textbf{b} Number of steps required for self-consistent (SC) and non-self-consistent (NSC) calculations. Box-plots show median as center line, upper and lower quartiles as box limits, 1.5x interquartile range as whiskers, and outliers as points. 
    \textbf{c} Non-self-consistent physical property calculations for total energy, atomic forces, band energies, and band gaps. Dashed lines indicate chemical accuracy (1 meV/atom for per-atom energies, 0.03 eV/\AA~ for forces, and 43 meV for band energies and band gap). Errors of exactly zero are replaced with $10^{-8}$ for visualization purposes. Interior of violin-plots shows median as center line, upper and lower quartiles as box limits, and 1.5x interquartile range as whiskers.
    \textbf{d} Percentage of materials achieving chemical accuracy for non-self-consistent calculations. 
    \label{fig:property-prediction-errors}
    }
\end{figure}

We demonstrate that charge densities predicted by \MODEL{} reduce the number of SCF steps, thereby accelerating DFT calculations, on unseen non-magnetic materials (Fig.~\ref{fig:property-prediction-errors}a, b) compared to a SAD initialization. 
The median number of SCF steps needed to converge a single-point calculation self-consistently for a non-magnetic material starting from the SAD initialization is 15, while the same starting from a \MODEL{} initialization is 11, representing a 26.7\% decrease in SCF steps. 
As an upper-bound baseline, we also include the distribution of the number of SCF steps needed to converge a calculation starting from a converged (i.e. ground-truth) charge density (SC), showing that our model still has room for improvement. 
About 96\% of calculations done with a predicted charge density ran with fewer SCF steps than initializing with SAD, and the compute time improvement is robust to a wide range of $\epsilon_{mae}$ (Supplementary Fig.~\ref{fig:scf_step_percent_savings}).

To evaluate how well \MODEL{}'s predicted charge densities reproduce physical properties, we compute the energy, forces, band energies, and bandgaps of the test set non-self-consistently (Fig.~\ref{fig:property-prediction-errors}c, d).
40\% of materials in the test set have per-atom energy errors less than 1 meV/atom. 
70\% of materials have forces less than 0.03 eV/\r{A}, a common threshold for structural relaxations; 15\% of the test set have force errors of 0.0 (Supplementary Fig.~\ref{fig:property_pair_correlation}). 
All of the structures with perfect agreement have self-consistent forces of 0.0 eV/\r{A} for each atom and Cartesian component, possibly because of cancellation of charge density errors due to symmetry. 
45\% and 76\% of materials recover band energies at each k-point and the direct band gap within chemical accuracy (43 meV), respectively. 
45\% of the band gaps obtained non-self-consistently exactly matched those of self-consistent calculations (Supplementary Fig.~\ref{fig:property_pair_correlation}). 
Similar to the force data described above, calculations with perfect agreement all have a self-consistent band gap of 0 eV, indicating the model is able to readily identify metallic materials.
We note that our use of chemical accuracy for band gaps is a more stringent criterion than experimental measurement errors for well-studied semiconductors~\cite{wingBandGapsCrystalline2021}. 
In addition, non-self-consistent calculations using the Materials Project parameters as done with this test set have an additional source of error (Supplementary Fig.~\ref{fig:nsc_error} and Methods) as a result of an incorrect calculation parameter set by the Materials Project.
This error could explain why not all NSC calculations computed on a self-consistent charge density achieve chemical accuracy (see Fig.~\ref{fig:property-prediction-errors}d and Fig. \ref{fig:sc-charge-density-lmaxmix-error}).

To examine generalization capabilities of \MODEL{} on out-of-distribution data, we analyze a random subset of 1924 materials published by GNoME \citep{merchant2023scaling} with 5 or more unique species to differentiate it from the MP distribution. 
The average $\epsilon_{mae}$ found for this set is 0.632\%, which is higher than that of the MP test set for \MODEL{} but lower than either DeepDFT variant trained on the Materials Project (Table \ref{table:nmape}). 
We conduct a similar experiment to the MP test set above with non-self-consistent DFT calculations on 1510 non-magnetic materials.
The median energy error is 4.3~meV/atom for the GNoME set, whereas the same is 1.8~meV/atom for the MP test set. 
Only 8.4\% of materials in GNoME computed non-self-consistently from the \MODEL{} charge density obtained the total energy within chemical accuracy (\textless1~meV/atom), though nearly all self-consistent calculations meet this criteria (Supplementary Fig.~\ref{fig:gnome_energy}).
The median number of SCF steps needed for self-consistency falls from 14 to 10 steps when initializing with \MODEL{}, a 28.6\% drop (Supplementary Fig.~\ref{fig:gnome_scf_steps}) and a similar acceleration found for the Materials Project test set. 
Moreover, 91\% of calculations saw a decreased number of SCF steps to reach convergence. 
While further work is needed to improve the generalizability of \MODEL{} for non-self-consistent property predictions, self-consistent calculations on out-of-distribution materials are still accelerated when initialized by the predicted charge density.

In addition to the model's robustness to materials outside its training set, we show that \MODEL{} can predict the charge densities of materials after small atomic perturbations with minor increase to the charge density error (Supplementary Fig.~\ref{fig:rattle}). 
This allows us to accurately obtain phonon properties, which may rely on finite differences as the main method to obtain approximations to the local potential energy surface, from non-self-consistent calculations. 
We compute the constant-volume heat capacity ($C_v$) under the harmonic approximation for two zeolites, which are not included in the Materials Project dataset, and show that $C_v(T)$ obtained closely match those obtained by self-consistent calculations (Supplementary Section \ref{zeolite-thermal}). 
As another example of transferability, we compute the electronic bandstructure of the Si(111)-2x1 surface non-self-consistently from the \MODEL{} prediction and show that it almost exactly matches that of a self-consistent calculation (Supplementary Section \ref{si-bs}). 

\subsection{Higher-order representations lead to improved model accuracy} \label{effects-of-order}

To demonstrate the impact of introducing higher-order equivariant tensor representations in our model, we examine the effect of training models while varying the maximum rotation order $L$, with $\ell \in \{0, ..., L\}$ up to $L=4$ used in the final model.
We determine the number of channels for each rotation order by $N_\text{channels} = \lfloor\frac{ 500 / (L + 1)}{\ell * 2 + 1}\rfloor$.
For example, the $L=0$ model has representations consisting of 500 even scalars,
and 500 odd scalars, while the $L=1$ model has representations consisting of 250
even scalars, 250 odd scalars, 83 even vectors, and 83 odd vectors.
In this way, each model is constructed such that the total representation size and the proportion of the total representation used by each rotation order
are approximately equal. We train each model on 1,000, and 10,000 material subsets and the full MP dataset, using the same validation and
test sets. We observe a consistent increase in
performance on each training subset as the maximum rotation order increases, and a similar scaling behavior with respect to the training set size (Supplementary Fig.~\ref{fig:nmape_lmax}). 
This trend suggests that higher-order representations are necessary for accurate charge
density modeling, and can match the performance of lower order models with less data.

\newcommand{\materiala}{\ch{Cs(H2PO4)}}
\newcommand{\materialb}{\ch{Rb2Sn6}}
\newcommand{\angularmetric}{\zeta}

\begin{figure}[t!]

  \centering
  \includegraphics[width=\linewidth]{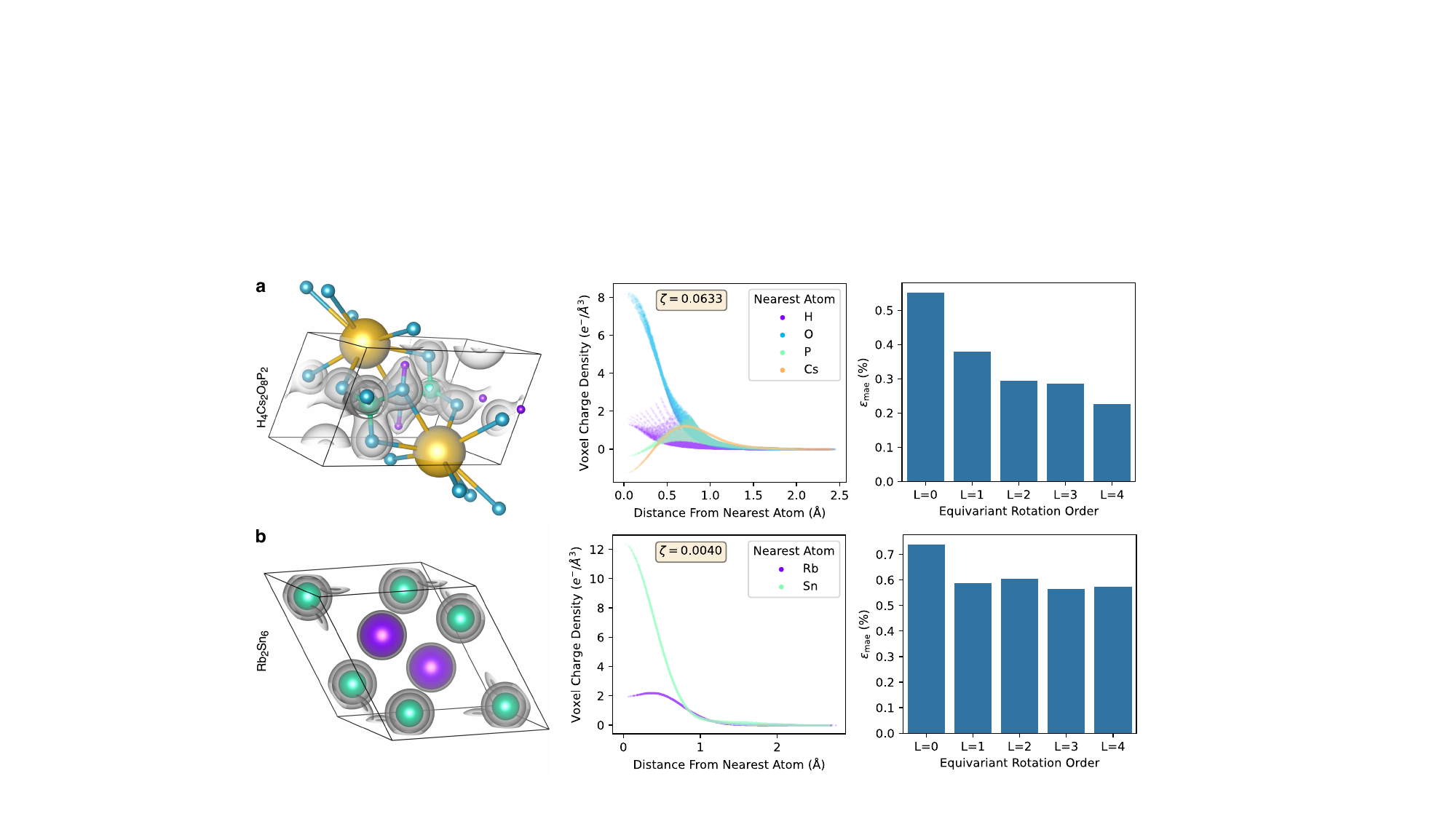}
  \caption{\textbf{Effect of angular variance.}
  Comparison of materials with high and low angular variance.
  \textbf{a} Material \materiala{} with high angular variance.  \textbf{b} Material \materialb{} with low angular variance. Left to right:
  visualization of charge density isosurfaces (gray);
  plot of DFT-computed charge density with respect to radial distance from nearest atom;
  $\epsilon_\text{mae}$ for \MODEL{} model predictions on these materials.}
  \label{fig:angular-variance-example}
\end{figure}

To gain intuition behind why and when higher-order representations yield higher performance, we investigate two factors contributing to the variance of charge density within a  material: (1) \textit{radial} dependence, or a dependence on the distance from neighboring atoms, and (2) \textit{angular} dependence, or dependence on angular orientation with respect to those atoms and the larger atomic system. While most materials exhibit strong radial dependence, some also exhibit strong angular dependence. For example, materials with covalent bonds share electrons between nuclei, leading to higher charge density along bonding axes and more angular dependence.
In Fig.~\ref{fig:angular-variance-example}, we illustrate this concept with two materials. \materiala{} exhibits high angular dependence, where charge density is dependent on angle as well as radial distance from the nearest atom. The complex nature of the charge density shown here arises from covalent bonding among the H, P, and O atoms.
Conversely, \materialb{} does not exhibit this, as its density appears
to be almost entirely a function of the atomic distance, suggesting that a
lower-order equivariant or invariant architecture could model its electron density well. This may be indicative of ionic interactions rather than the covalent interactions found in \materiala{}.
We find this intuition to be consistent with model performance, as
Fig. \ref{fig:angular-variance-example} shows similar performance for $L=0$ and $L=4$ models
for \materialb{}, while the $L=4$ model exceeds the performance of the $L=0$ model for
\materiala{}.

\begin{figure}
    \centering
    \includegraphics[width=\linewidth]{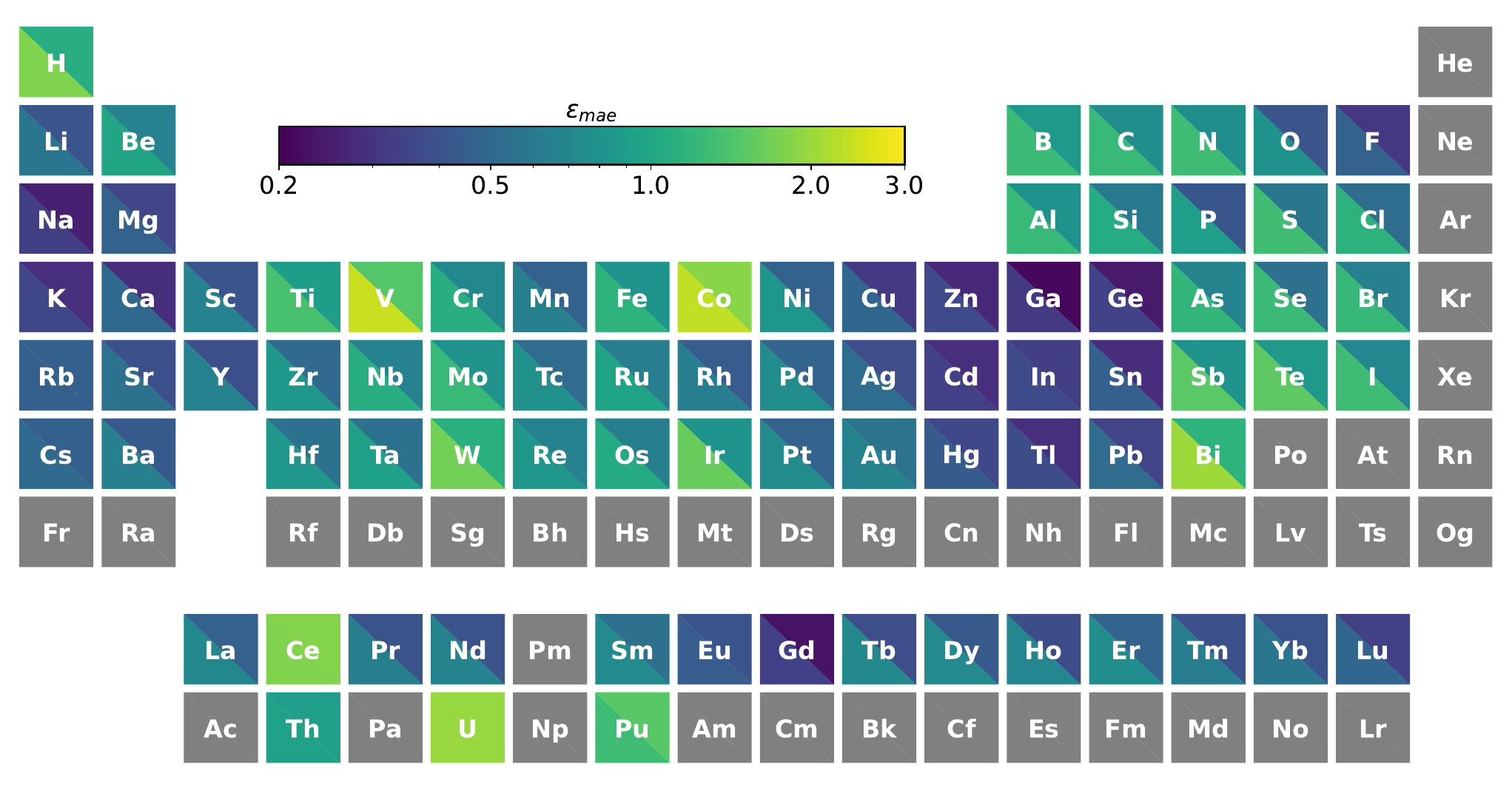}
    \caption{\textbf{Effect of rotation order on element-wise error.}
    The lower-left triangles indicate the average $\epsilon_\text{mae}$ per element at $L=0$, and the upper right triangles represent $\epsilon_\text{mae}$ per element at $L=4$. Performance improvements are shown across the periodic table. Values are computed over the set of grid points $G_\text{local}$,
    and averaged across all materials containing those elements in the test set. Only elements in at least 10 test set materials are shown.}
    \label{fig:periodic_L0_L4}
\end{figure}

At the atomic level, we find more evidence of the benefits of higher-order equivariant representations in modeling charge density. We examine the per-element errors by computing $\epsilon_\text{mae}$ on grid points $G_{\text{local}, i} \subset G$, a set of points that are within 2\AA{} of an atom of species $i$.
Fig.~\ref{fig:periodic_L0_L4} shows this per-element error for both $L=0$ (lower-left triangles) and $L=4$ (upper-right triangles), averaged across the test set. While the $L=0$ model struggles with reactive non-metals and metalloids, which often form covalent bonds, $L=4$ shows strong improvement among these elements. Nearly all elements show improvement, but the median change in $\epsilon_{mae}$ values for those materials containing a non-metal or metalloid is $-44.6\%$ ($n=1628$) compared to $-23.0\%$ for materials with only metals ($n=346$) (Supplementary Fig.~\ref{fig:metal_violin}).

To further quantify angular dependence within the system, we develop a metric $\angularmetric{}$ to determine to what extent
a particular system exhibits more angular variation in its charge density with respect to atom
locations. This is achieved by measuring the dot product of the unit vector from a probe point
to its nearest neighboring atom and the gradient of the density at that probe point:

\begin{equation}
  \angularmetric(G) = 1 - \frac
  {\sum_{\vec{g}_k \in G}|\nabla \rho(\vec{g}_k) \cdot \hat{r}_{ki}|}
  {\sum_{\vec{g}_k \in G}|| \nabla \rho(\vec{g}_k) ||}
\end{equation}

where $G$ is a set of probe points and $\hat{r}_{ki}$ is a unit vector from probe
point at location $\vec{g}_k$ to the closest atom at location $\vec{r}_i$.
Intuitively, a material with charge density gradients pointing directly towards or away from the nearest atom will have dot products larger in magnitude and $\angularmetric\rightarrow0$, whereas a material with density that has high angular variance with respect to the nearest atom will have dot products smaller in magnitude, and $\angularmetric\rightarrow1$; see Supplementary Fig.~\ref{fig:periodic_zeta} for the test set average $\angularmetric$ over $G_{local,i}$ for every element. We observe that the differences in performance between the lower and higher rotation order networks correlates strongly with $\angularmetric$ (Supplementary Fig.~\ref{fig:angular-variance-scatter}). 
With covalent bonding and associated high angular variance frequently present in the Materials Project data,
this further justifies the need for introducing higher rotation order representations
into charge density prediction networks. In addition, we find the higher-order model maintains higher accuracy under perturbations to atomic positions (Supplementary Fig.~\ref{fig:rattle}), an imperative trait when using models for phonon calculations.

\subsection{Runtime and scalability}

To demonstrate the scalability of our model, we analyze the runtime duration of
our model on systems with increasing number of atoms, and compare to the
duration of DFT calculations. We run each method on a single material, \ch{BaVS3} (mp-3451 in
MP), creating supercells from $1\times 1 \times 1$ to $10\times 10 \times 10$
and recording the runtime to generate charge density of the system. Figure
\ref{fig:runtime} shows an approximate linear, $O(N)$, scaling of our model with respect
to number of atoms, while DFT exhibits approximately $ O(N^{2.3})$ before exceeding our
computational budget. This is to be expected, as the graph size increases
linearly with cell volume if the resolution remains the same, while DFT has
shown to scale at a cubic rate with respect to system size~\citep{jorgensen2022equivariant}.
Furthermore, our method can be fully parallelized up to each point in the density grid with no communication overhead.

\section{Discussion}

This work introduces \MODEL{}, an architecture for predicting charge density through equivariance with higher-order representations. 
We demonstrate that introducing higher-order tensor representations over scalars and vectors used in prior work achieves greater predictive accuracy, and show how this is achieved through the improved modelling of systems with high angular variance.
\MODEL{} enables accurate charge density calculations on atomic systems larger than what is computationally feasible with DFT, but can also be used in combination with self-consistent DFT to reduce the number of SCF steps needed for convergence. 
Our findings hold across the large and diverse Materials Project dataset, paving the way towards a universal charge density model for materials.  Future work to expand the model to charge density calculations of unrelaxed structures such as those in MPtrj \citep{dengCHGNetPretrainedUniversal2023} would support general application of ML models to accelerate ab initio calculations and should be pursued.

The work of Hohenberg and Kohn proves that all material properties could, in principle, be derived from the electron density~\cite{PhysRev.136.B864}. 
Because this quantity is so fundamental to predicting material properties, we view models that compute electron densities as a potential intermediate towards other material properties that may improve the generalizability of machine learning models in material science. As evidence, our non-self-consistent calculations, which can be seen as a ground-truth calculation of the electronic strucuture (and derived properties) using the crystal structure and charge density as inputs, demonstrate that one could indeed obtain electronic and thermal properties of materials using highly accurate charge-density estimates. 
On a broader dataset, we show that non-self-consistent calculations are able to capture the total energy, atomic forces, electronic bandstructure, and the band gap within chemical accuracy for a majority of structures in the test set, though there is room for improvement. This leads us to believe that large models trained on a charge density dataset that spans the known material space could be a promising direction towards a foundation model for DFT calculations that can be fine-tuned for downstream property prediction.

We have also explored the connection between errors in charge density predictions and the property error. We show in Supplementary Fig. \ref{fig:per_atom_energy_error} - \ref{fig:gap_error_zero_hist} that errors in the charge density establish a lower bound to the property error, and that small charge density errors can give rise to substantial property error. Examining charge density errors more closely, specifically analyzing $\epsilon_{mae}$ localization in low- and high-error non-self-consistent calculations, we find difficulty identifying localized regions where charge density prediction errors lead to property errors (Supplementary Fig. \ref{fig:material_distance_error}). 
While the errors in non-self-consistent property predictions are certainly affected by errors in charge density predictions from \MODEL{}, the errors may in part be explained by an incorrect DFT setting (the VASP tag LMAXMIX. See the Methods section for more information).
The errors introduced by this calculation parameter may mean achieving chemical accuracy is improbable for some materials using non-self-consistent calculations even when initializing with a self-consistent (i.e. ground truth) charge density.

We demonstrate higher-order E(3) equivariant tensor operations yield significant improvement in charge density predictions and are more data efficient (Fig.~\ref{fig:angular-variance-example} and \ref{fig:periodic_L0_L4}). However, the E(3) convolution used in this work are limited by the $O(L^6)$ time complexity of tensor products~\cite{passaro2023reducing}. 
Recent work~\citep{passaro2023reducing} has showed that SO(3) convolutions can be reduced to SO(2), with a time complexity of $O(L^3)$. In the future, our approach can be combined with the methodology presented in~\citep{passaro2023reducing} to improve computational efficiency, and enable the practical usage of higher rotation orders ($L>>4$) in atomistic modeling. 

We also anticipate extending our model to include the spin density, which will allow for acceleration of DFT calculations of magnetic materials. 
Training ChargE3Net with energies and forces would allow for rapid identification of the lowest energy magnetic configuration, a normally laborious and time-intensive process \cite{hortonHighthroughputPredictionGroundstate2019,leeIdentificationGroundstateSpin2017}, and the ability to obtain magnetic properties, such as the Curie temperature.

\section{Methods}

\subsection{\MODEL{} architecture} \label{architecture}

As described in Section \ref{equivariance}, the inputs of the model consist of probe coordinates $\vec{g}$ and atom coordinates $\vec{r}$ with species $z$. Each atom and probe point are represented by a dictionary of tensors, where $A^n$ denotes an atom representation and $P^n$ denotes a probe representation, each at layer $n$.
 $A^{n=0}$ is initialized as a one-hot encoded $z$, represented as a $\ell=0$, $p=1$ tensor with $N_\text{channels}$ equal
to the number of atomic species.
$P^{n=0}$ is initialized as single scalar zero, with $\ell=0$, $p=1$, and $N_\text{channels}=1$.
Each representation is updated through a
series of alternating convolution $\text{Conv}(\cdot)$ and non-linearity
$\text{Gate}(\cdot)$ layers, a total of $N_\text{interactions}=3$ times. With each convolution layer, all possible outputs of the tensor product at different $\ell$ and $p$ values with $\ell_o \leq L$ are concatenated, increasing the representation size past the scalar initializations of $A^n$ and $P^n$.
Atom representations are updated with $A_i^{n+1} =
\text{Gate}(\text{Conv}^n_\text{atom}(\vec{r}_i, A_i^n))$.
$\text{Conv}_\text{atom}$ is defined as:

\begin{equation}\label{eqn:convolution}
  \text{Conv}^n_\text{atom}(\vec{r}_{i}, A^n_i) =  W^n_1 \left( \sum_{j\in N(i)} W^n_2 A^n_j \otimes R(r_{ij}) Y(\hat{r}_{ij}) \right) + W^n_3 A^n_i
\end{equation}

where $W_1, W_2, W_3$ are learned weights applied as a linear mix or self-interaction
\citep{thomas2018tensor}. The set $N(i)$ includes all atoms within the cutoff
distance, including those outside potential periodic boundaries.     
$r_{ij}$ is the distance from $\vec{r}_i$ and $\vec{r}_j$, with unit vector $\hat{r}_{ij}$.
$Y(\hat{r}_{ij})$ are spherical harmonics, and $R(r_{ij}) \in
\mathbb{R}^{N_\text{basis}}$ is a learned radial basis function defined as:

\begin{equation}
  R(r_{ij}) = \text{MLP}([\phi_1(r_{ij}), ..., \phi_{N_\text{basis}}(r_{ij})])
\end{equation}

where $\text{MLP}(\cdot)$ is a two layer multilayer perceptron with SiLU non-linearity~\cite{hendrycks2016gaussian}, $\phi(\cdot)$ is a Gaussian basis
function $\phi(r_{ij})_k \propto \exp(-(r_{ij} - \mu_k)^2)$ with $\mu_k$
uniformly sampled between zero and the cutoff, then normalized to have a zero
mean and unit variance.

The convolution updates the representation of each atom to be the sum of tensor products between neighboring atom representations and the corresponding spherical harmonics describing their relative angles, weighted by a learned radial
representation. This sum is followed by an additional self-connection, then a residual self-connection.
The output of the convolution is then passed though an equivariant 
non-linearity $\text{Gate}(\cdot)$ operation as described in \cite{weiler20183d}. We use
SiLU and $\tanh$ activation functions for even and odd parity scalars respectively, as is 
done in \cite{batzner20223}.

Probe representations are updated similarly as the atoms for each layer, except their representations depend
solely on the representations of neighboring atoms, with no probe-probe interaction. Each probe representation is
updated with
$P_k^{n+1} = \text{Gate}(\text{Conv}^n_\text{probe}(\vec{g}_k, P_k^n))$, where
$\text{Conv}_\text{probe}$ is defined as:

\begin{equation}
  \text{Conv}^n_\text{probe}(\vec{g}_{k}, P^n_k) =  W^n_1 \left( \sum_{i\in N(k)} W^n_2 A^n_i \otimes R(r_{ik}) Y(\hat{r}_{ik}) \right) + W^n_3 P^n_k\hspace{0.1in}.
\end{equation}

Note that weights $W$ are not shared with those for $\text{Conv}^n_\text{atom}(\cdot)$. Since
atom representations are computed independently of probe positions, they can be
computed once per atomic configuration, even if multiple batches are required to
obtain predictions for all probe points. Finally, predicted charge density $\hat{\rho}(\vec{g}_k)$ is computed as 
a linear combination of the scalar elements of the final representation $P_k^{n = N_\text{layers}}$.

\subsection{Training \MODEL{} models}

We develop an algorithm to achieve more efficient atom-probe graph construction, enabling faster training and inference times. Prior work constructs graphs by adding probe points directly to the atomic structure, computing edges with atomic simulation software such as \texttt{ase} or \texttt{asap} \citep{larsen2017atomic}, then removing edges between probes. While this is performant for small systems with few probes, we find it does not scale well due to the redundant computation of probe-probe edges, which grow quadratically with respect to the number of probe points. To address this, our algorithm leverages two \textit{k}-d trees \citep{bentley1975multidimensional}; one to partition atoms, and one to partition probe points. This enables an efficient radius search for atom-probe edges exclusively. We implement this through modification of the \textit{k}-d tree implementation \texttt{scipy} \citep{2020SciPy-NMeth} to handle periodic boundary conditions. Constructing graphs with 1000 probe points, we observe an average graph construction time ($\pm$ standard error) of $0.00758\pm0.00087$, $0.0103\pm0.0005$, and $0.00462\pm0.00056$ seconds for MP, QM9, and NMC respectively. This is at least an order of magnitude faster than methods from \citep{jorgensen2022equivariant}, with which we measured graph construction times of $12.6\pm0.9$, $0.114\pm0.006$, and $0.141\pm0.010$ seconds.

To train the \MODEL{} model, Supplementary Table \ref{table:setup} outlines the experimental setup on each of the datasets (e.g., learning rate, batch size, number of probe points and number of training steps). Specifically, at each gradient step, a random batch of materials is
selected, from which a random subset of $N_\text{points}$ probe points are selected. We use
the Adam optimizer \citep{kingma2014adam}, and decay the learning rate by
$0.96^{s/\beta}$ at step $s$. We find that optimizing for L1 error improves
performance and training stability over mean squared error (Supplementary Table \ref{table:l1}). Supplementary Fig.~\ref{fig:loss} shows the evolution of training loss and validation error as the model is trained on each dataset. All models were
implemented in PyTorch and trained on NVIDIA Volta V100 GPUs using a distributed compute architecture.

\subsection{Datasets} \label{datasets}

The QM9 and NMC datasets, along with train, validation and test splits, are provided by
\citet{jorgensen2022equivariant}. The QM9 dataset\citep{qm9vasp} contains VASP calculations of the 133,845 small organic molecules introduced by \citet{ramakrishnan2014quantum}, with training, validation, and test splits of size 123835, 50, and 10,000 respectively. The NMC dataset\citep{nmcvasp} consists of VASP calculations of 2000 randomly sampled nickel manganese cobalt oxides containing varying levels of lithium content, with training, validation, and test splits of size 1450, 50, and 500 respectively.
For the MP data, we collected 122,689 structures and
associated charge densities from
\url{api.materialsproject.org} \cite{shen2022representation}.
Structures in the dataset that shared composition and space group were identified as duplicates and only the highest \texttt{material\_id} structure was included, leaving 108,683 materials. The data was split randomly into training, validation, and
test splits with sizes 106,171, 512, and 2000 respectively. Later, analyzing the test set materials for thermodynamic stability and structural abnormalities (Supplementary Fig.~\ref{fig:energy_volume}), 26 materials were found to have unrealistic volume per atom, in excess of $100$ \r{A}$^3$/atom and mostly from space group \textit{Immm}. These materials were removed from the test set and excluded from all reported results and figures.

We note to the reader a known issue with the Materials Project dataset. Original calculation parameters may have an incorrectly set VASP parameter called LMAXMIX, which sets the maximum angular momentum number for the one-center PAW charge densities saved in the charge density output of VASP and passed into the charge density mixer. This affects elements with d- and f-electrons when no Hubbard U parameter was set by the Materials Project. In Supplementary Fig. \ref{fig:nsc_error}, we correct this error for the randomly sampled test set, then compared the distribution of properties between non-self-consistent calculations of the original and corrected input parameters. Note that non-self-consistent calculations with a potentially incorrectly set LMAXMIX, as is set in the current Materials Project charge density dataset, makes a per-atom energy error about two orders of magnitude larger than that of self-consistent calculations. This represents an additional source of error to all non-self-consistent calculations based on Materials Project data.

In addition to a held-out test set randomly sampled from the Materials Project, we obtain a random set of 2000 published structures from GNoME \citep{merchant2023scaling} as a second test set. Specifically, we choose a random set of materials from GNoME with 5 or 6 unique species as these are less well-represented in the Materials Project training data. DFT calculations were set up using pymatgen \cite{ong2013python} and otherwise using the same methodology as detailed in Section \ref{dft-methods}. 76 calculations failed, leaving 1924 materials for analysis, 1510 of which were non-magnetic. Non-magnetic structures were identified via DFT calculations initialized by SAD and magnetic moment guesses provided by pymatgen with an overall absolute magnetic moment \textless 0.1 $\mu B$ and absolute atomic magnetic moments \textless 0.1 $\mu B$ for every atom in the structure.

Supplementary Fig. \ref{fig:tsne} shows a visualization of SOAP descriptors of materials in the MP train and test set and GNoME subset, projected into 2-d using t-SNE. We do not find a significant difference in distributions across the MP train and test sets, which is to be expected as the data is divided randomly. Figure~\ref{fig:sunburst} shows the distribution of crystal systems and space groups within the MP training set. As seen with the SOAP descriptors, in Fig.~\ref{fig:radarplot} we note a similar distribution of crystal systems across the MP train and test sets, with the GNoME subset showing significantly more monoclinic systems and significantly fewer tetragonal and cubic systems.

\subsection{Density Functional Theory calculations}
\label{dft-methods}

All density functional theory (DFT) calculations were done with VASP version 5.4.4 \cite{hafner2008ab,kresse1993initio,kresse1996efficiency,kresse1996efficient} using projector-augmented wave (PAW) pseudopotentials \cite{kresse1999ultrasoft}.  
For all calculations done on the test set of Materials Project data, we use the input files provided by the Materials Project to ensure these are computed in a manner consistent with how the charge density grids are originally generated. 
For the GNoME dataset, we use a k-point density of 100/\r{A}\textsuperscript{-3}, the same density that the original Materials Project charge density dataset was computed with, and generated the input files with pymatgen~\cite{ong2013python}.
We set LMAXMIX to 4 for d-elements and 6 for f-elements only for DFT+U calculations as set by pymatgen, which, while incorrect (see Section \ref{datasets}), allows us to compare our performance on this set of materials against that of the Materials Project test set.
For the experiments with perturbed structures, including those where electronic and thermal properties were computed, we use all the same settings as the Materials Project but with different k-grids, all generated with the Monkhorst-Pack scheme unless otherwise specified. 
We use a 520 eV energy cutoff and converge each calculation until the change in energy between SCF steps is less than $10^-5$ eV. 
For the Si(111)-(2x1) calculation, we use a k-grid of 4x8x1, while all zeolite calculations were done at the gamma-point only.
Electronic bandstructures were plotted using pymatgen \cite{ong2013python}. 
We use the same Hubbard U values as the Materials Project, which were obtained from Wang et al. \cite{wang2006oxidation} and Jain et al. \cite{jain2011formation}.
PAW augmentation occupancies were not predicted by \MODEL{} and were taken from the Materials Project dataset or computed self-consistently for the GNoME dataset.

Thermal properties were computed using phonopy \cite{phonopy-phono3py-JPCM,phonopy-phono3py-JPSJ} for the heat capacity. Zeolite and bulk Si structures were relaxed until the maximum norm of the Hellmann-Feynman forces fell below 0.001 eV/\r{A} prior to computing thermal properties. 
Structures were perturbed by 0.03\r{A} to find the Hessian of the potential energy surface via finite differences, and the heat capacity at constant volume can then be found via the harmonic approximation with the equation

\begin{equation}
    C_v(T) = R\sum_k \exp \left( \theta_k / T \right) \left( \frac{\theta_k/T}{\exp\left( \theta_k/T \right) - 1} \right)
\end{equation}

where $\theta_k=h\nu_k/k_B$, $h$ is Planck's constant, $\nu_k$ is the vibrational frequency of vibrational mode $k$, $k_B$ is Boltzmann's constant,  $R$ is the gas constant, and $T$ is the temperature in Kelvin.

\section{Data availability}

Materials Project data was collected 7 May 2023. Associated task identifiers will be included in our provided repository, along with the necessary download script, and our train, validation, and test splits.
We also provide the identifiers for the subset of GNoME data used our work. 
VASP computations on the QM9 and NMC datasets were downloaded from \citet{qm9vasp} and \citet{nmcvasp}  respectively.

\section{Code availability}

Our model implementation, along with pretrained model weights is available at \url{https://github.com/AIforGreatGood/charge3net}.

\section*{Declarations}
\subsection*{Acknowledgements}

We thank Tess Smidt and Mark Polking for helpful discussions.

DISTRIBUTION STATEMENT A. Approved for public release. Distribution is unlimited. This material is based upon work supported by the Under Secretary of Defense for Research and Engineering under Air Force Contract No. FA8702-15-D-0001. Any opinions, findings, conclusions or recommendations expressed in this material are those of the author(s) and do not necessarily reflect the views of the Under Secretary of Defense for Research and Engineering. © 2023 Massachusetts Institute of Technology. Delivered to the U.S. Government with Unlimited Rights, as defined in DFARS Part 252.227-7013 or 7014 (Feb 2014). Notwithstanding any copyright notice, U.S. Government rights in this work are defined by DFARS 252.227-7013 or DFARS 252.227-7014 as detailed above. Use of this work other than as specifically authorized by the U.S. Government may violate any copyrights that exist in this work.

\subsection*{Author Contributions}

T.K. designed, implemented, and trained the \MODEL{} models. K.Q. collected and prepared the MP, QM9, and NMC datasets, and contributed to the software implementation and training of baseline models. K.Q. devised the angular variation metric and performed analysis on the effects of rotation order together with T.K. E.T. performed experimentation and analysis for initialization of DFT calculations and non-self-consistent calculation of material properties. K.T. provided guidance on DFT calculations. L.L. conceived the project, supervised the research and advised experimentation and data analysis. All authors contributed to the writing and editing of the manuscript.

\subsection*{Competing interests}
All authors T.K., K.Q., E.T., K.T. and L.L. declare no financial and non-financial competing interests.

\bibliography{refs}

\appendix
\counterwithin{figure}{section}
\counterwithin{table}{section}

\newpage
\section{Supplementary Information}

\subsection{Additional training details}
\label{additional-training-details}

For training the invDeepDFT and equiDeepDFT models on the MP dataset, we use the same training and model configurations as the original work, training for $10^6$ steps. For NMC and QM9 datasets, we use the authors released models, which are reported to have been trained for $5\times10^6$ and $6.5\times10^6$ steps respectively. We note that this is comparable to the number of steps \MODEL{} is trained for on these datasets, as shown in Table \ref{table:l1}.

\subsection{Heat capacity and thermal conductivity of faujasite}
\label{zeolite-thermal}

\begin{figure}
    \centering
    \includegraphics[width=\linewidth]{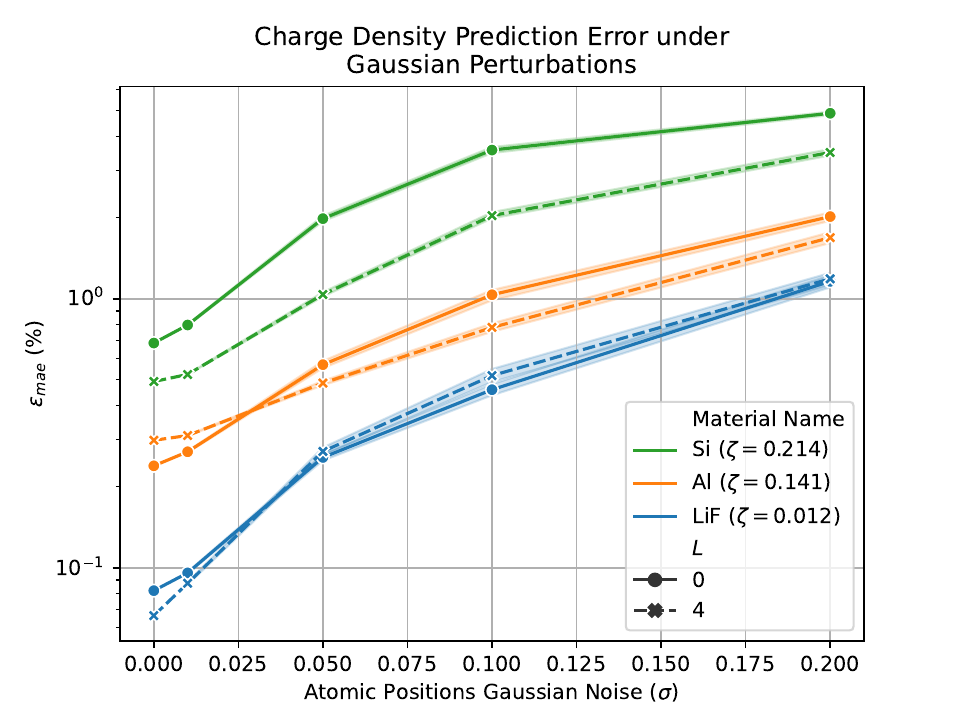}
    \caption{$\epsilon_{mae}$ as a function of perturbations of atomic positions for LiF (mp-1138, blue), Al (mp-134, orange), and Si (mp-149, green). Solid lines represent the $L=0$ variant of \MODEL{}, while dashed lines represent the $L=4$ variant of \MODEL{}. We compute self-consistent DFT calculations with atomic positions $r'=r+\delta$, where $r$ is the original position of the atoms from the Materials Project and $\delta\sim\mathcal{N}(0, \sigma)$. These are used as the ground truth charge densities which we use to compare with \MODEL{}'s predictions. $\angularmetric$ values are computed for the unperturbed structure charge density. Shaded regions depict 95\% CI.}
    \label{fig:rattle}
\end{figure}

Computing thermal properties of materials require knowing the phonon modes of the material. 
A typical approach to finding phonon modes and their corresponding vibrational frequencies is to first compute the Hessian of the total energy with respect to the atomic positions, yielding a second-order (harmonic) approximation to the potential energy surface. 
The Hessian can be found via finite differences, and the number of displacements required scales with the size of the system under study. 
These computationally intensive calculations may be made cheaper with accurate charge density estimates from \MODEL{}, and we show that the charge density predictions are robust to small perturbations of atomic positions (see Figure \ref{fig:rattle}).
Below, we further show that even non-self-consistent calculations of thermal properties with these estimates yield near-perfect agreement with self-consistent calculations of the same. 

\begin{figure}
    \centering
    \begin{subfigure}[b]{0.4\textwidth}
        \centering
        \includegraphics[width=\textwidth,trim={24cm 0 24cm 0},clip]{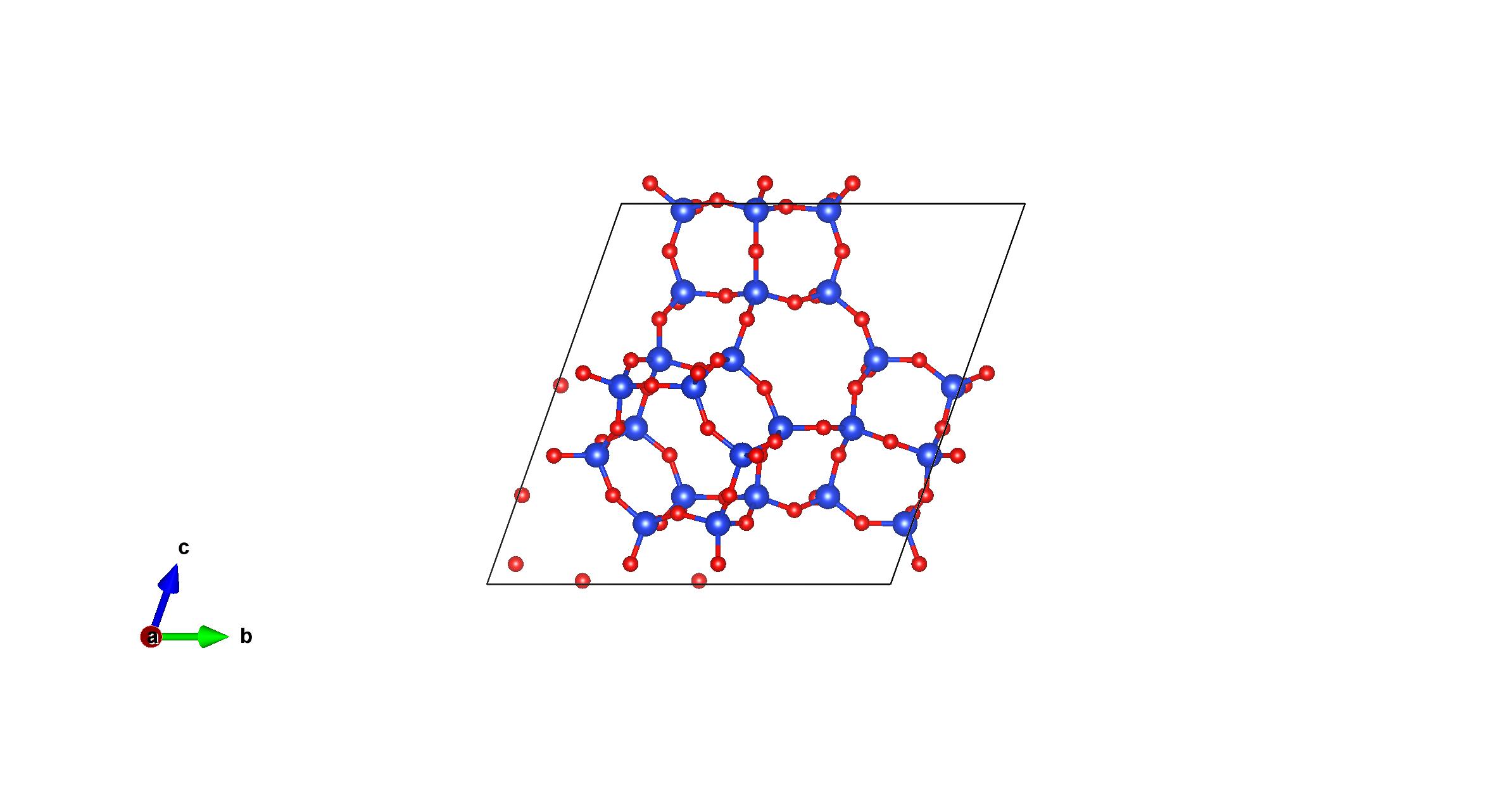}
    \end{subfigure}
    \begin{subfigure}[b]{0.4\textwidth}
        \centering
        \includegraphics[width=\textwidth,trim={24cm 0 24cm 0},clip]{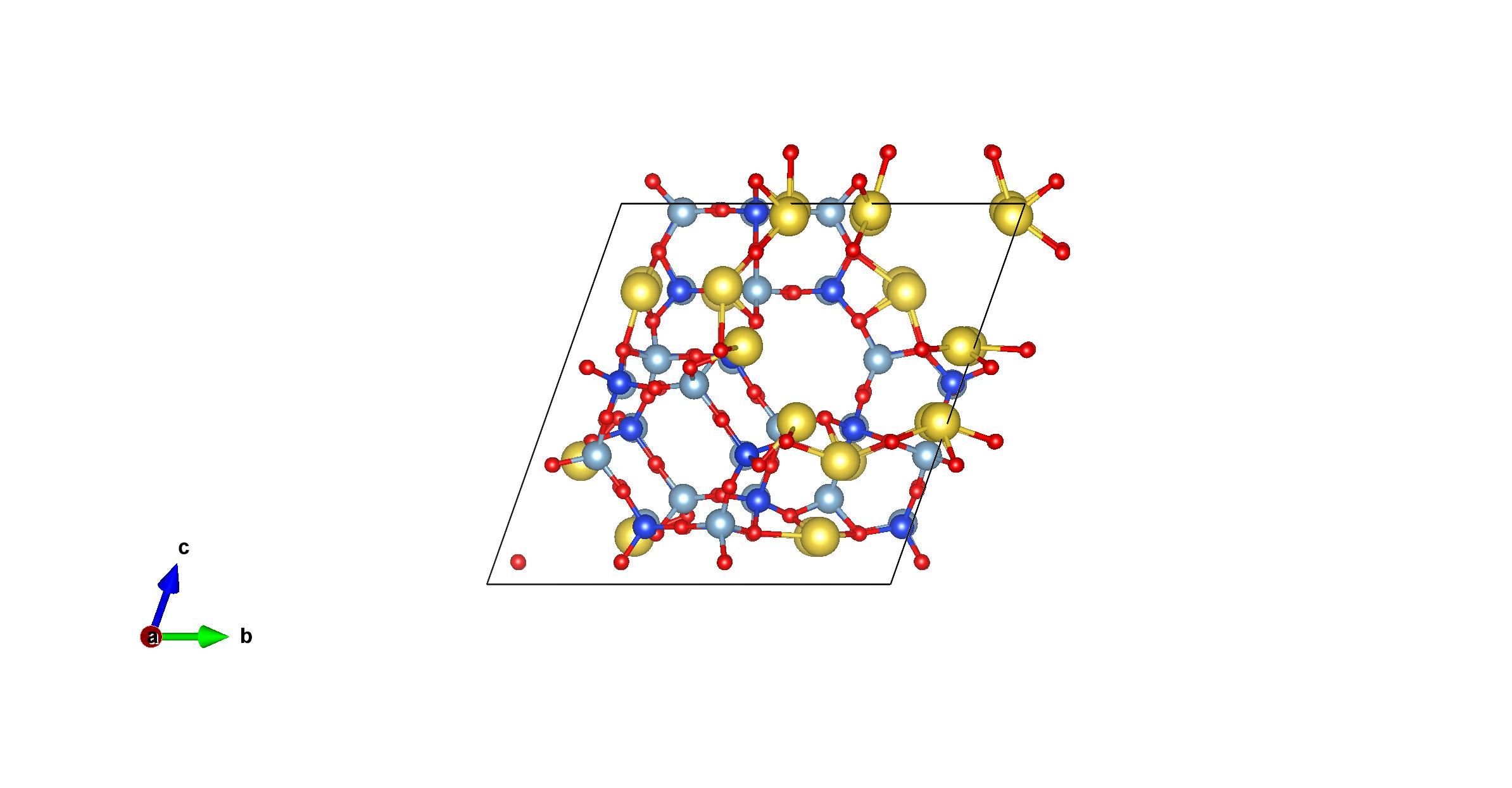}
    \end{subfigure}
    \caption{Structural models of (right) USY and (left) NaX zeolites, two types of faujasite zeolites. Blue = Si, red = O, grey = Al, yellow = Na.}
    \label{fig:faujasite-model}
\end{figure}

\begin{figure}
    \centering
    \includegraphics[width=0.75\linewidth]{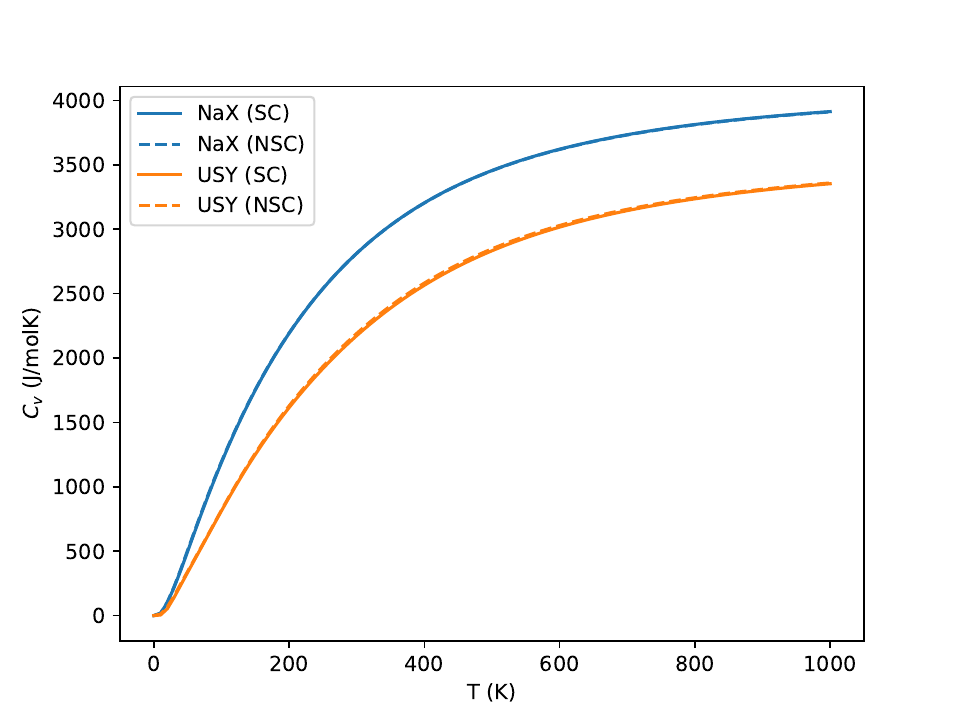}
    \caption{Constant-volume heat capacity as a function of temperature as computed from self-consistent DFT calculations and non-self-consistent calculations on the predicted charge densities. Blue curves are for NaX and orange curves are for USY. Solid curves are computed via self-consistent DFT calculations of atomic displacements, while dashed curves are computed via non-self-consistent DFT calculations using \MODEL{}'s predicted charge densities of the same.}
    \label{fig:faujasite-thermal}
\end{figure}

We evaluate two variants of faujasite zeolite, one that is purely composed of Si and O (USY) \cite{parise1984structure}, and another where a portion of the Si sites are exchanged with Al charge-balanced by Na ions (NaX). 
The structures of these are shown in \ref{fig:faujasite-model}.
Understanding the thermal properties of porous materials such as these via data-driven predictive models could aid high-throughput screening of gas adsorbents for industrially relevant separation processes such as carbon capture \cite{nikolaidis2018modelbased,young2023processinformed,moosavi2022datascience}. 
We note that while the elemental compositions of USY and NaX (Si-O and Na-Al-Si-O, respectively) exist in the Materials Project dataset, porous zeolites are not found in the dataset. 
Despite this, we show that the constant-volume heat capacity ($C_v$) obtained via the harmonic approximation from non-self-consistent calculations on \MODEL{}'s predicted charge densities closely resemble the $C_v$ obtained by self-consistent DFT (see Figure \ref{fig:faujasite-thermal}). 
Charge density predictions of USY and NaX had an average $\epsilon_\text{mae}$ of 0.16\% and 0.58\%, respectively. 
This demonstrates \MODEL{}'s ability to obtain accurate macroscopic properties of unseen structures by predicting local perturbations to the charge density 

\subsection{Electronic bandstructure of Si surface}
\label{si-bs}

DFT calculations on material surfaces are highly important for understanding catalytic reactions, material interfaces, and corrosion \cite{sholl2009density} Because surfaces are not part of the training set, obtaining accurate charge densities from \MODEL{} shows one example of the model's ability to generalize to unseen atomic environments. 

\begin{figure}
    \centering
    \begin{subfigure}[b]{0.4\textwidth}
        \centering
        \includegraphics[width=\textwidth,trim={35cm 0 35cm 0},clip]{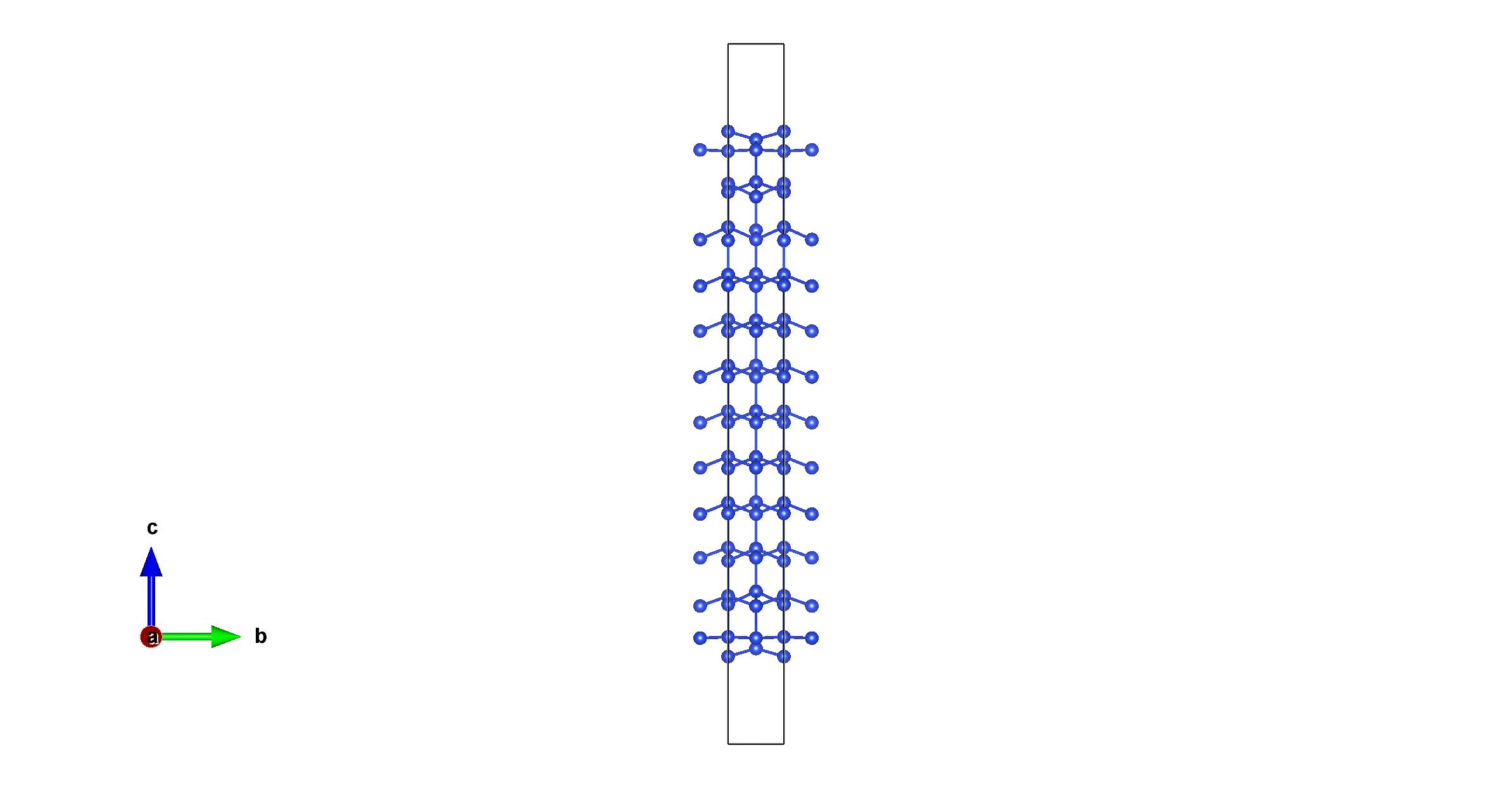}
    \end{subfigure}
    \begin{subfigure}[b]{0.4\textwidth}
        \centering
        \includegraphics[width=\textwidth,trim={35cm 0 35cm 0},clip]{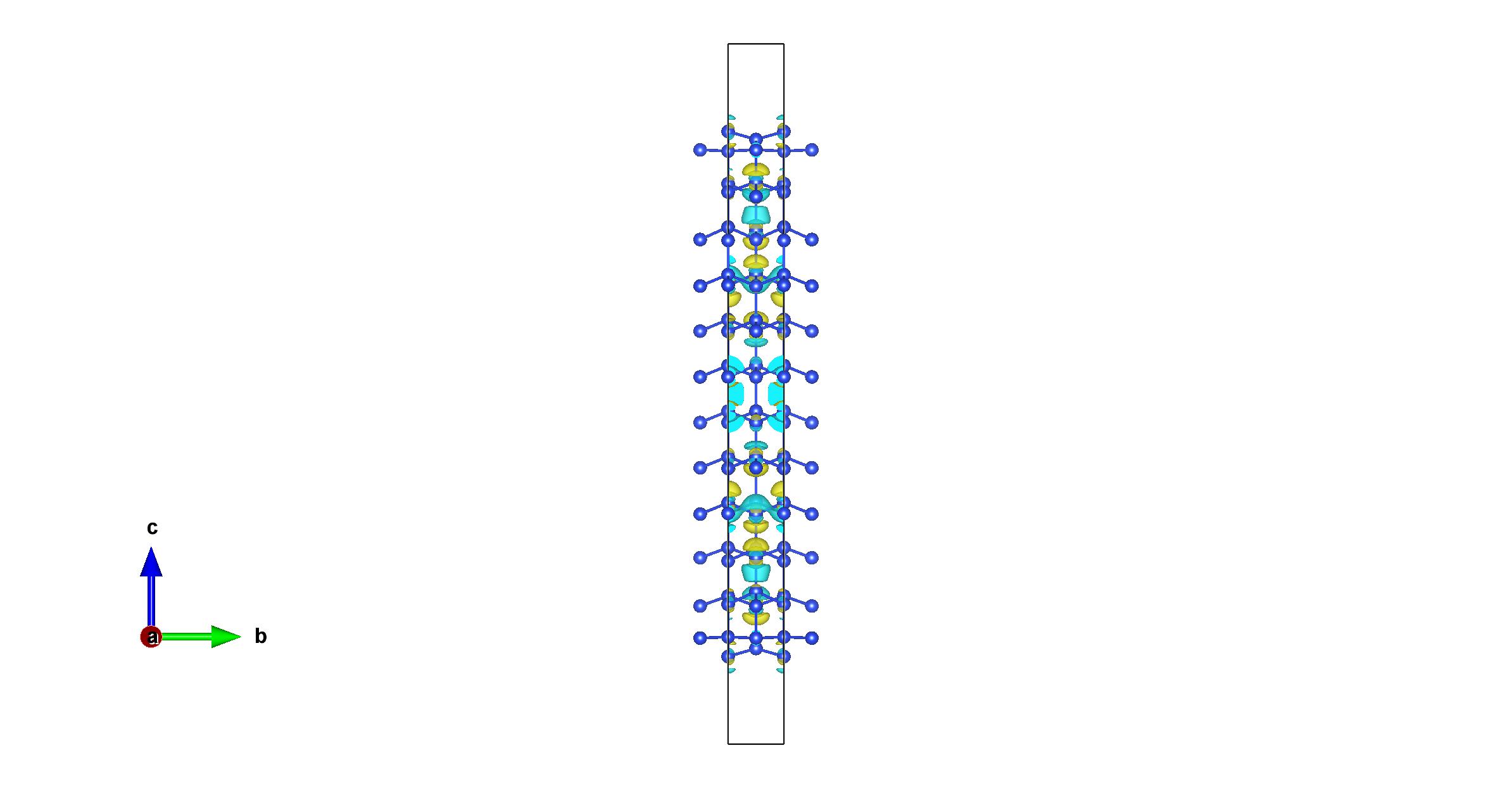}
    \end{subfigure}
    \caption{(right) Structural model of a Si(111)-(2x1) surface, and (left) an isosurface of the charge density error overlayed on the structural model. Yellow denotes overestimation while teal denotes underestimation. The isosurface level used here is approximately $1.6\times 10^{-6}$ e/\AA$^3$. }
    \label{fig:si-surface-model}
\end{figure}

\begin{figure}
    \centering
    \includegraphics[width=0.75\linewidth]{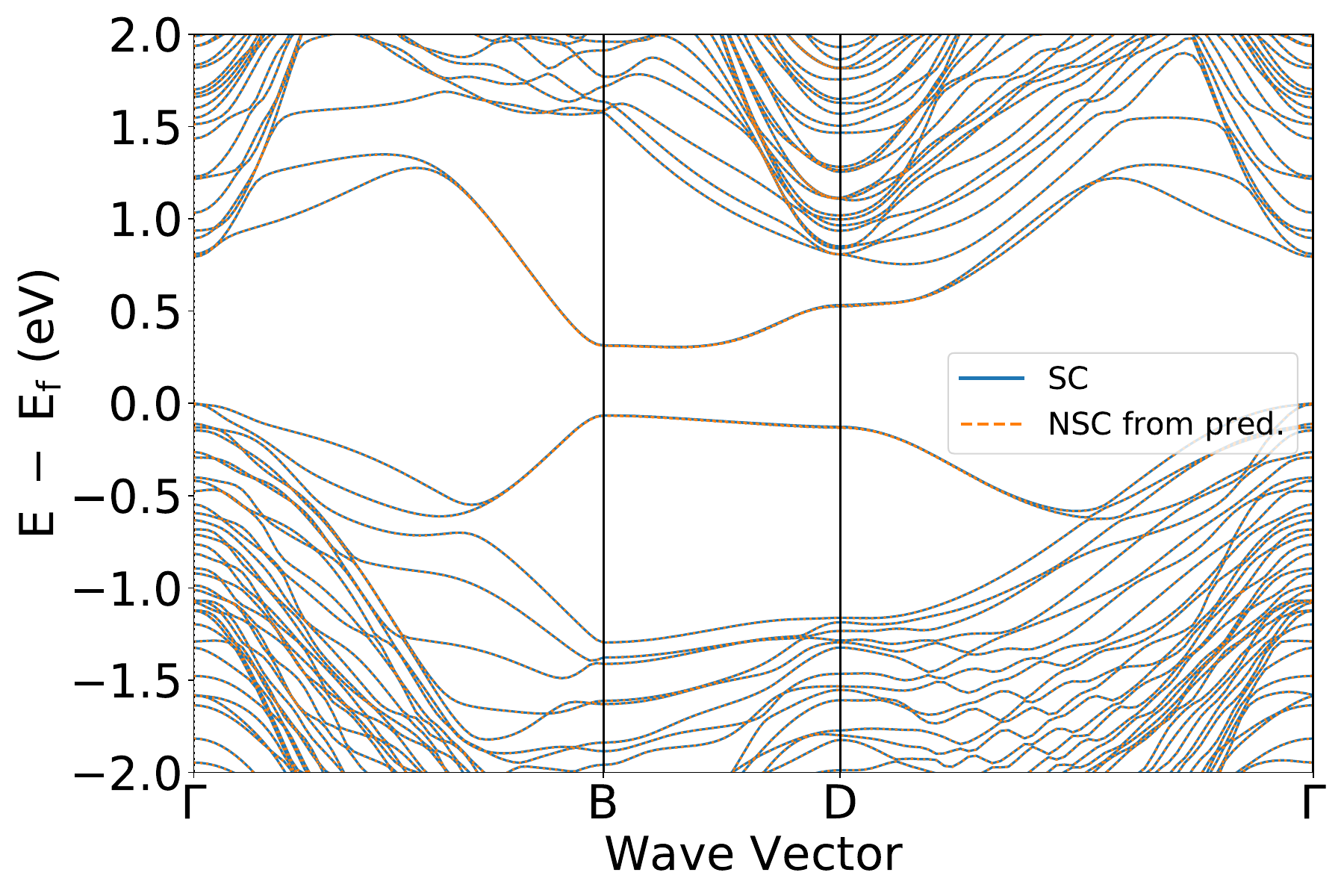}
    \caption{Electronic bandstructure of Si(111)-(2x1) surface. Blue represents a self-consistent (SC) DFT calculation with the PBE exchange-correlation functional, and orange represents a non-self-consistent (NSC) calculation using the charge density predicted by \MODEL{} with the same functional.}
    \label{fig:si-surface-bs}
\end{figure}

Pandey shows through tight-binding calculations that the Si(111)-(2x1) surface, the structure of which is plotted in \ref{fig:si-surface-model}, is characterized by pairs of Si sites that undergo significant bond rearrangement to form zig-zag patterns of raised Si atoms. 
These pairs form chains of $\pi$-bonds across the surface. 
While multiple polymorphs of Si, both experimentally-observed and hypothesized, are catalogued in the Materials Project and part of the training set, \MODEL{} was not trained on any Si surface charge densities.
We obtain the structure of the surface model from \cite{jainReliabilityHybridFunctionals2011}, and in Figure \ref{fig:si-surface-bs}, we show that a non-self-consistent (NSC) calculation using the charge density output by \MODEL{} reproduces almost exactly the electronic bandstructure of a self-consistent (SC) DFT calculation. 
Interestingly, even the bands corresponding to the surface Si sites at the valence band maximum and conduction band minimum are accurately reproduced, indicating the model is capable of extrapolating the charge density grid in a physically meaningful way despite new atomic environments such as 3-coordinate surface Si sites. 
The $\epsilon_\text{mae}$ of the predicted charge density is approximately 0.83\%, and the distribution of charge density prediction error is shown in \ref{fig:si-surface-model}.

\newpage

\begin{table}[!h]
  \caption{Training setup}
  \label{table:setup}
  \centering
  \begin{tabular}{lllllll}
    \toprule
    Dataset           & Learning Rate & Decay $\beta$ &  $L$ & Batch Size & $N_\text{points}$ & Steps \\
    \midrule
    NMC               & 0.01  & $10^4$ & 4 & 8 & 200 & $7.5 * 10^5$\\
    QM9               & 0.01  & $10^4$ & 4 & 8 & 200 & $10^6$\\
    MP                & 0.005 & $3 * 10^3$ & 4 & 16 & 200 & $10^6$ \\
    \bottomrule
  \end{tabular}
\end{table}

\begin{table}[!h]
  \caption{Loss function ablation, reported in average $\epsilon_\text{mae}$ (\%), $\pm$
  one standard error.}
  \label{table:l1}
  \centering
  \begin{tabular}{llll}
    \toprule
    Loss Function     & MP                       & QM9                      & NMC \\
    \midrule
    MSE               & $0.722\pm0.011$          & $0.268\pm0.001$          & $0.076\pm0.001$\\
    L1                & $\mathbf{0.523}\pm0.010$ & $\mathbf{0.196}\pm0.001$ & $\mathbf{0.060}\pm0.001$ \\
    \bottomrule
  \end{tabular}
\end{table}

\newpage

\begin{figure}[!h]
    \centering
  \includegraphics[width=0.75\linewidth]{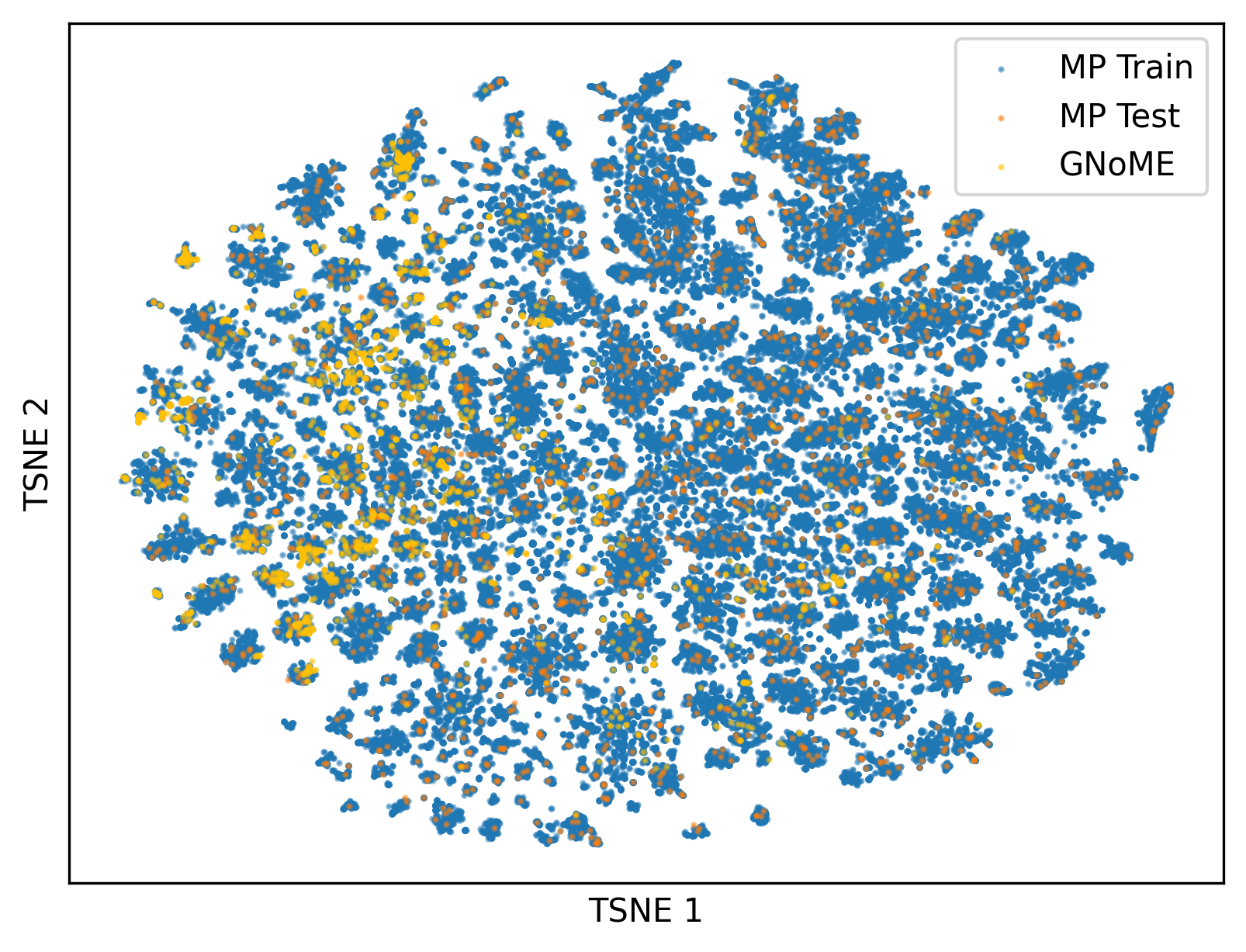}
  \caption{t-SNE visualization of Smooth Overlap of Atomic Positions (SOAP) descriptors of the MP train and test subsets, along the GNoME subset used for evaluation. Descriptors are computed using DScribe \cite{dscribe} with a cutoff radius of 4 \AA~and compressed \cite{darby2022compressing} with $\mu=1,\nu=1$. }
  \label{fig:tsne}
\end{figure}

\begin{figure}
    \centering
    \includegraphics[width=0.75\linewidth]{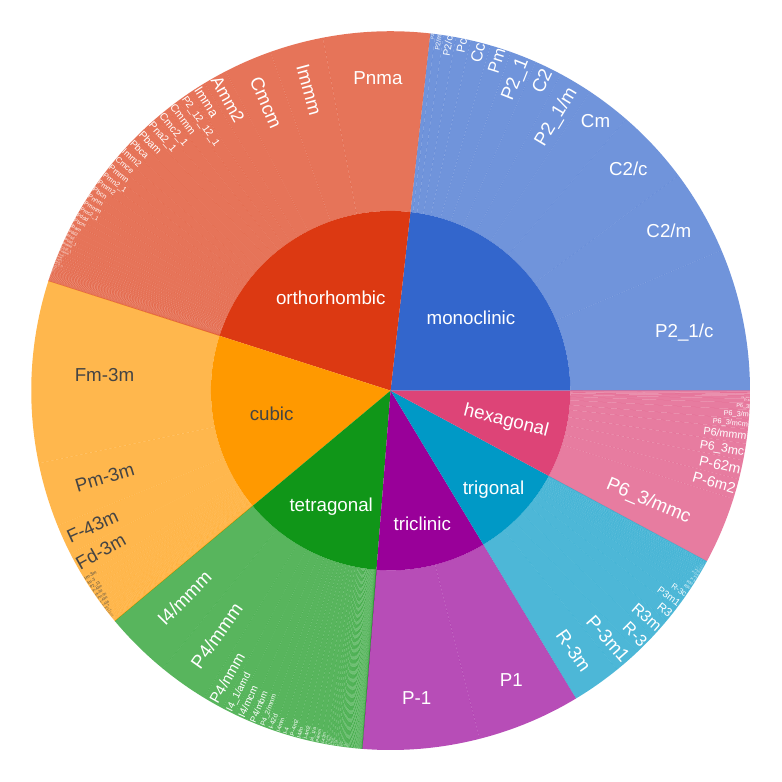}
    \caption{Sunburst plot showing the Materials Project training split crystal systems and space groups. Created using PyMatViz \citep{riebesell_pymatviz_2022}}
    \label{fig:sunburst}
\end{figure}

\begin{figure}
    \centering
    \includegraphics[width=0.75\linewidth]{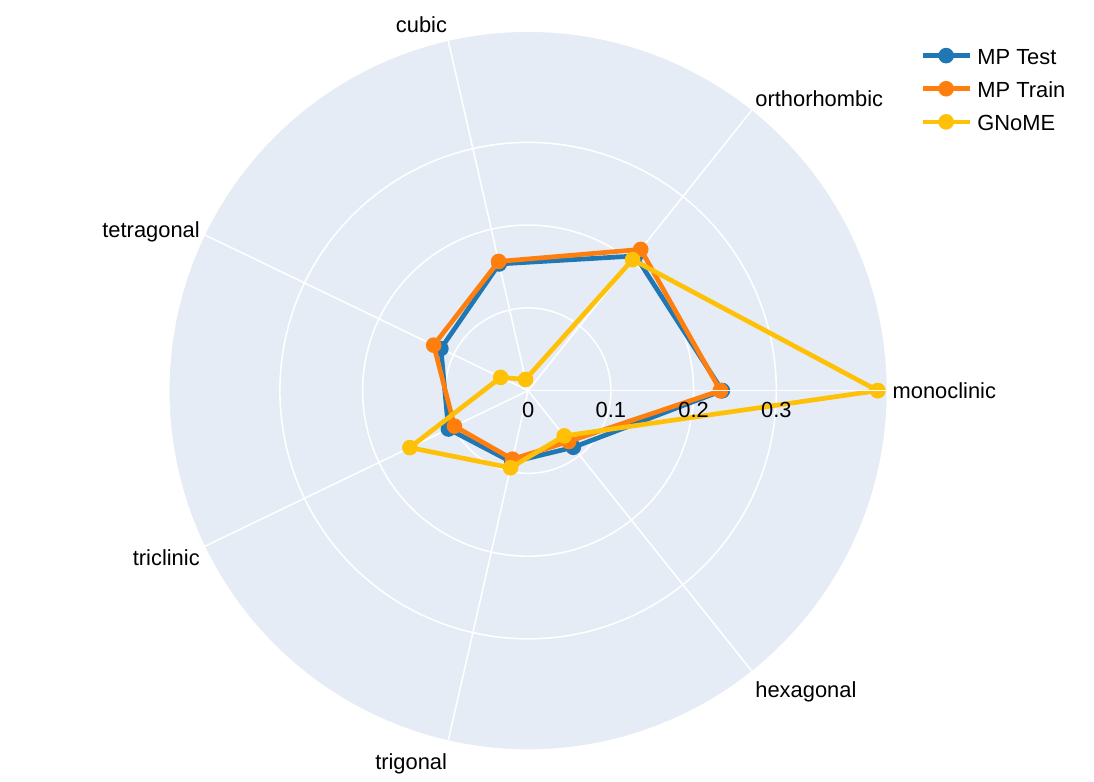}
    \caption{Radarplot showing crystal system distribution in Materials Project and GNoME subset. The GNoME subset is made up of materials with five or more unique atomic species. Markers indicate the fraction of each subset which has a particular crystal system structure.}
    \label{fig:radarplot}
\end{figure}

\begin{figure}
    \centering
    \includegraphics[width=0.85\linewidth]{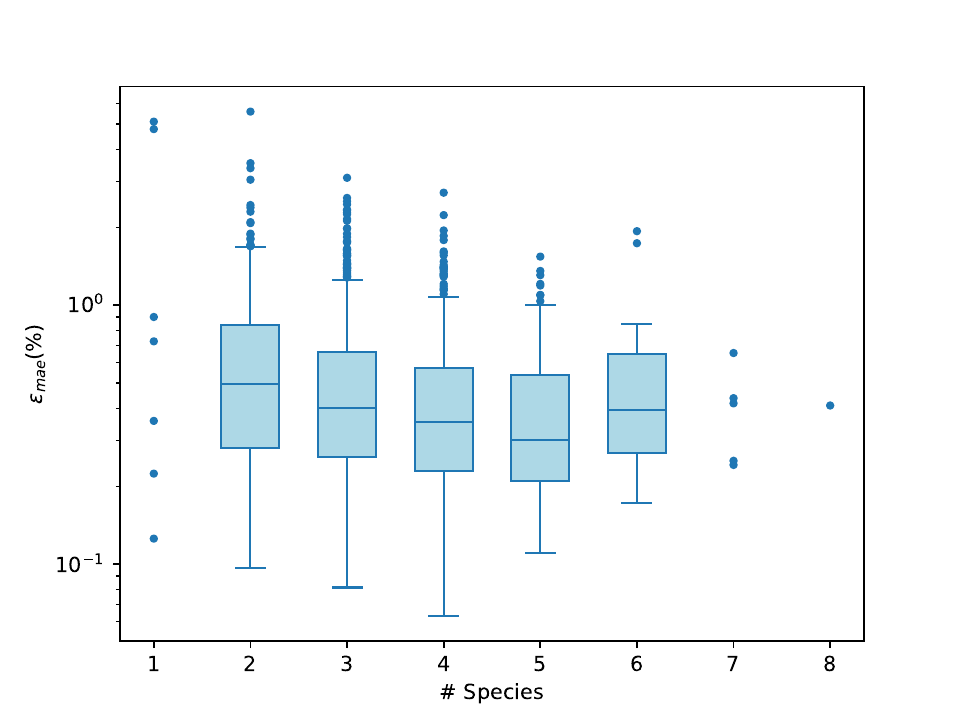}
    \caption{Box-plot showing $\epsilon_{mae}$ vs. number of species (unique elements) in Materials Project test set materials. Boxes shown for levels with at least 10 materials in the test set.  Box-plots show median as center line, upper and lower quartiles as box limits, 1.5x interquartile range as whiskers, and outliers as points.}
    \label{fig:nmape_species}
\end{figure}

\begin{figure}
    \centering
    \includegraphics[width=\linewidth]{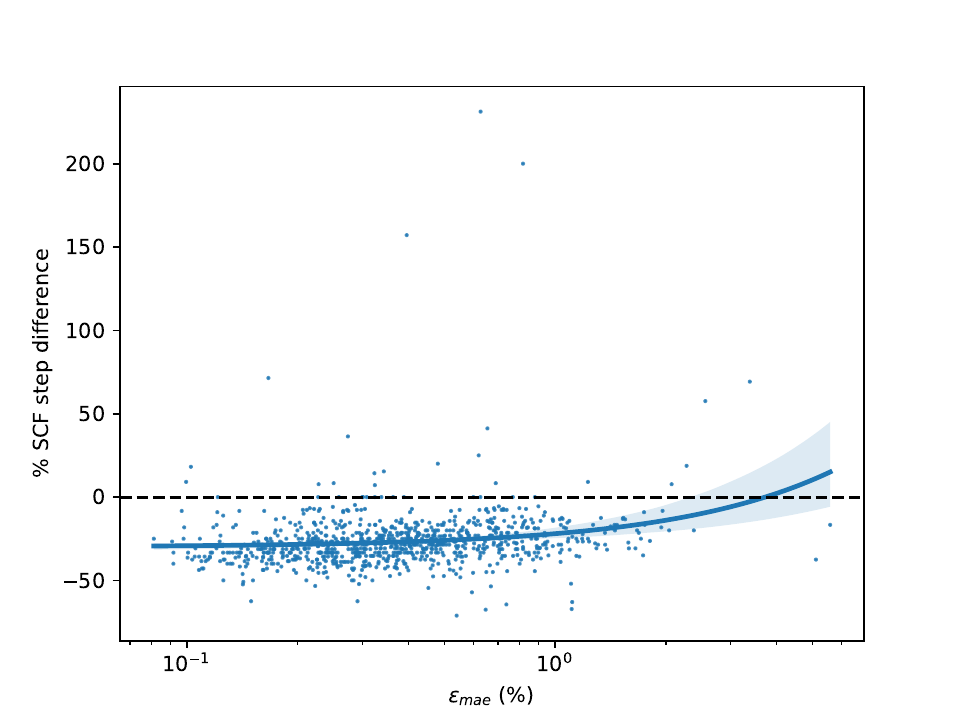}
    \caption{$\epsilon_{mae}$ plotted against the percent improvement in the number of SCF steps needed to reach self-consistency. Pearson $r\approx 0.19$ for the correlation between $\epsilon_{mae}$ and \% SCF step difference. Regression line is shown with 95\% CI shaded}
    \label{fig:scf_step_percent_savings}
\end{figure}

\begin{figure}
    \centering
    \includegraphics[width=\linewidth]{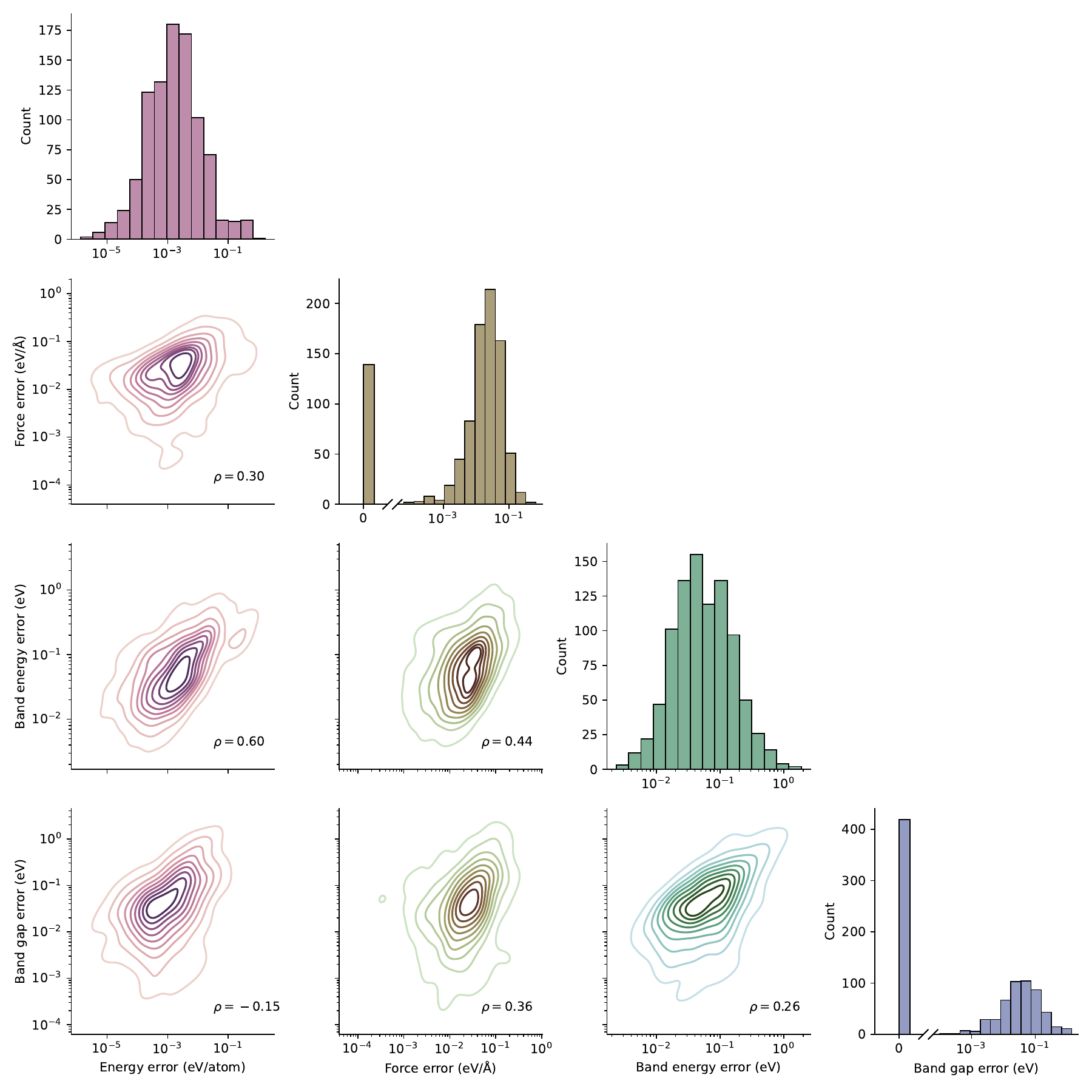}
    \caption{(diagonal) Distribution of property errors computed from a non-self-consistent calculation with a \MODEL{} charge density initialization, and (off-diagonal) density of correlations between property errors. $\rho$ denotes Spearman's correlation coefficients for each off-diagonal plot.}
    \label{fig:property_pair_correlation}
\end{figure}

\begin{figure}
    \centering
    \includegraphics[height=0.7\textheight]{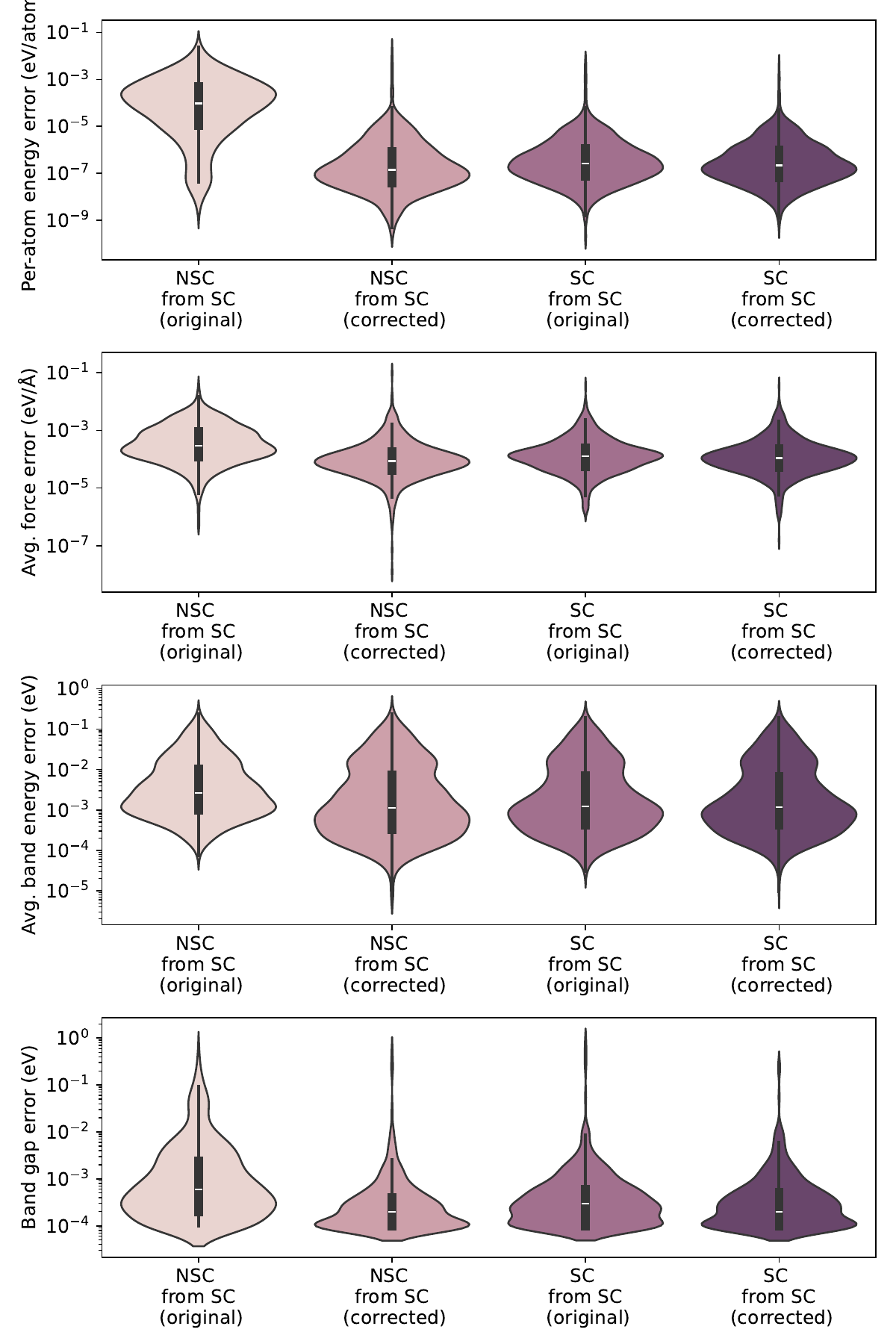}
    \caption{Charge densities of the Materials Project test set were computed self-consistently, then computed non-self-consistently (NSC) and self-consistently (SC). (original) denotes the LMAXMIX parameter originally used by Materials Project. (corrected) sets the LMAXMIX parameter to 4 for d-elements and 6 for f-elements. Errors of each property are calculated with respect to the initial self-consistent calculation initialized from SAD. Interior of violin-plots shows median as center line, upper and lower quartiles as box limits, and 1.5x interquartile range as whiskers}
    \label{fig:nsc_error}
\end{figure}

\begin{figure}
    \centering
    \includegraphics[width=\linewidth]{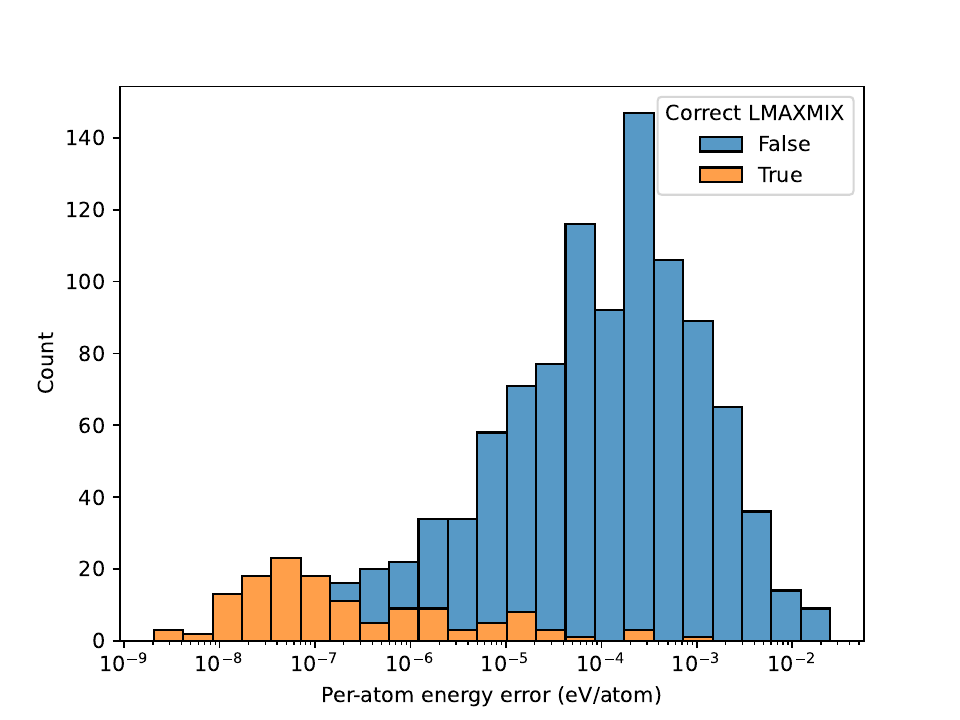}
    \caption{Per-atom energy error of non-self-consistent calculations done with self-consistent charge densities using inputs provided by the Materials Project for the test set used in this study. The Materials Project inputs have the LMAXMIX parameter correctly set for only DFT+U calculations (LMAXMIX = 4 for d-elements and = 6 for f-elements).}
    \label{fig:sc-charge-density-lmaxmix-error}
\end{figure}

\begin{figure}
    \centering
    \includegraphics[width=\linewidth]{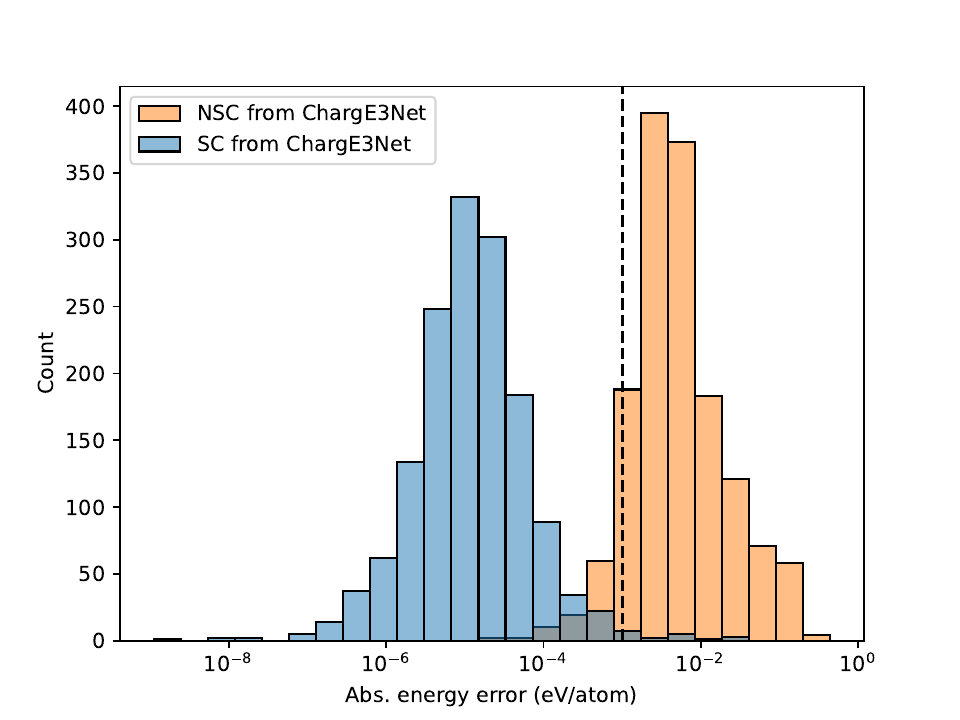}
    \caption{Per-atom total energy error of self-consistent (SC, blue) and non-self-consistent (NSC, orange) calculations initialized with \MODEL{} on a subset of GNoME materials. Dashed vertical line denotes chemical accuracy at 1 meV/atom.}
    \label{fig:gnome_energy}
\end{figure}

\begin{figure}
    \centering
    \includegraphics[width=\linewidth]{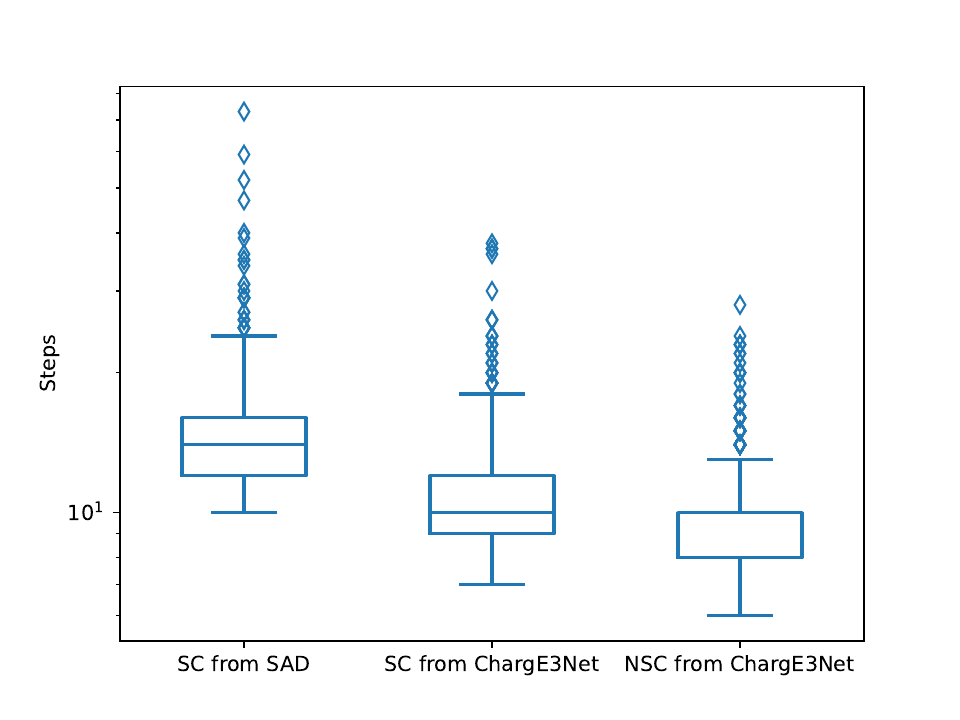}
    \caption{Distributions of SCF steps required for convergence of a subset of GNoME materials. Box-plots show median as center line, upper and lower quartiles as box limits, 1.5x interquartile range as whiskers, and outliers as points. The median for NSC from \MODEL{} is 8 and overlayed on the 25th percentile mark.}
    \label{fig:gnome_scf_steps}
\end{figure}

\begin{figure}[!h]
  \centering
  \includegraphics[width=0.9\linewidth]{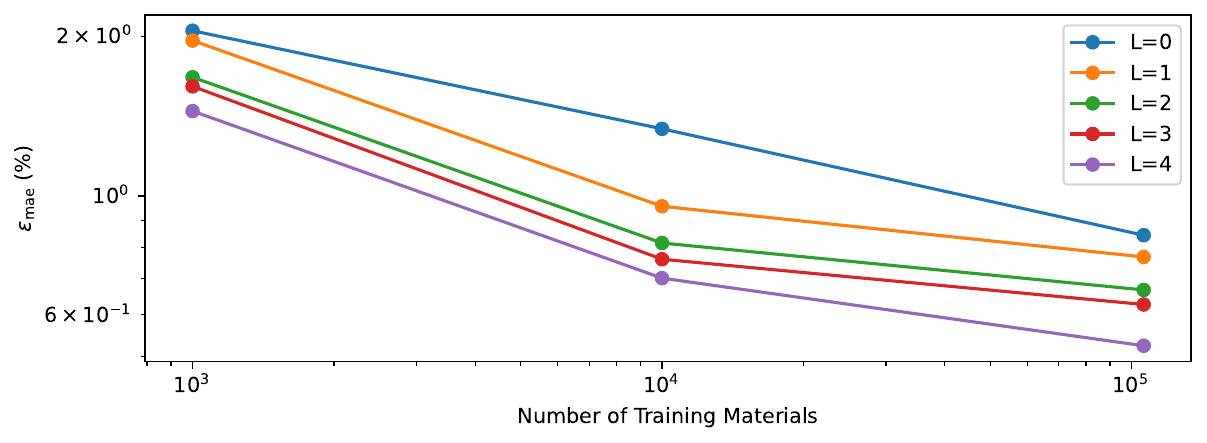}
  \caption{
    Log-log plot of training set size vs performance, measured in average
    $\epsilon_\text{mae}$ (\%) on the Materials Project test set. We show models train with
    maximum rotation order $L \in {0, 1, 2, 3, 4}$.}
  \label{fig:nmape_lmax}
\end{figure}

\begin{figure}[h]
    \centering
    \includegraphics[width=0.5\linewidth]{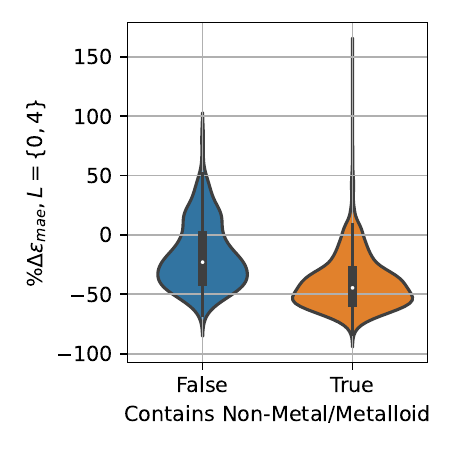}
    \caption{Violin plot of the percent change in $\epsilon_{mae}$(\%) from the $L=0$ to $L=4$ model for test set materials. 1648 materials contain either a non-metal or a metalloid (orange), while 346 materials are composed of only metals (blue). Interior of violin-plots shows median as center line, upper and lower quartiles as box limits, and 1.5x interquartile range as whiskers}
    \label{fig:metal_violin}
\end{figure}

\begin{figure}[h]
    \centering
    \includegraphics[width=\linewidth]{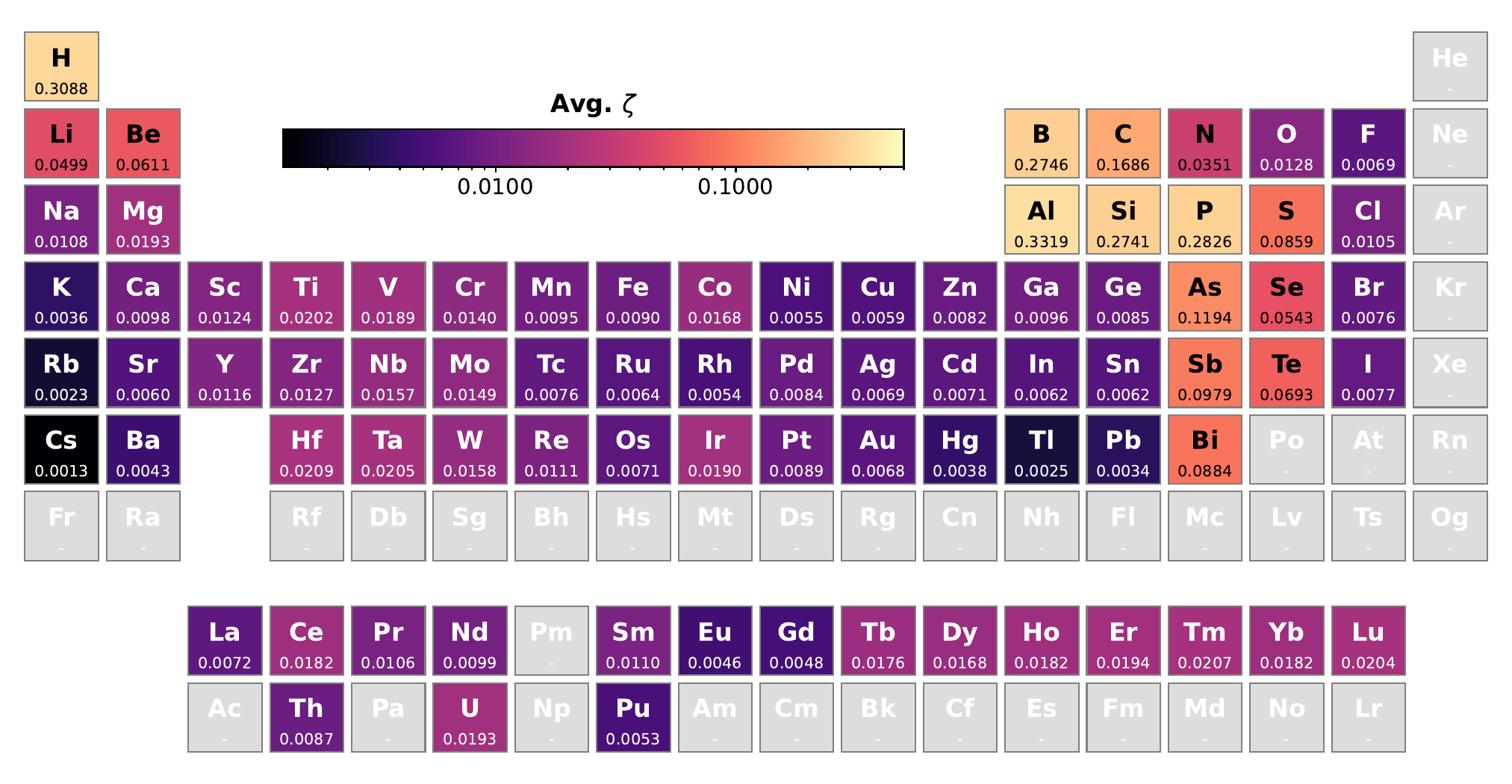}
    \caption{Average $\zeta$, computed on DFT-computed charge density at probe points $G_{local,i}$ for each element. Only elements contained in at least 10 test set materials shown.}
    \label{fig:periodic_zeta}
\end{figure}

\begin{figure}
    \centering
    \begin{subfigure}[b]{0.49\textwidth}
        \centering
        \includegraphics[width=\textwidth]{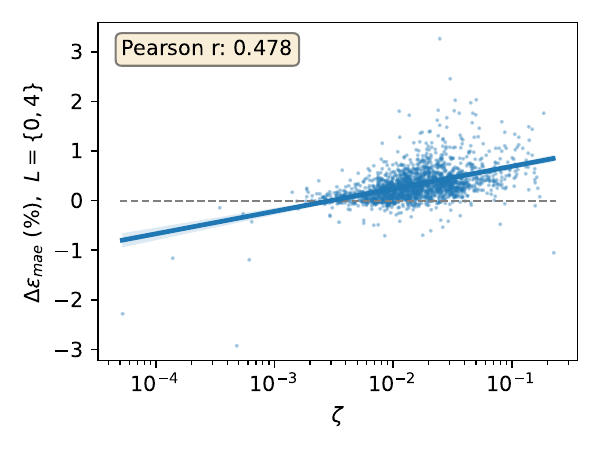}
        \caption{Per-Material ($n=1974$)}
    \end{subfigure}
    \hfill
    \begin{subfigure}[b]{0.49\textwidth}  
        \centering 
        \includegraphics[width=\textwidth]{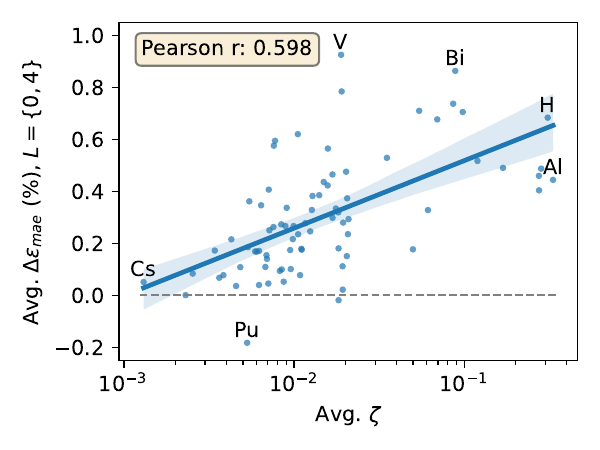}
        \caption{Per-Element ($n=85$)}
    \end{subfigure}
    \caption{$\Delta\epsilon_{mae}$ (improvement from $L=0$ to $L=4$) vs. $\zeta$, on a per-material (left) and per-element (right) basis. For per-element results, the $\epsilon_{mae}$ and $\zeta$ values are computed over the set of grid points $G_{local}$, and averaged across all materials containing that element. Only those elements in at least 10 test set materials are shown. Regression line is shown with 95\% CI shaded region for both charts.}
    \label{fig:angular-variance-scatter}
\end{figure}

\begin{figure}[h]
    \centering
    \includegraphics[width=\linewidth]{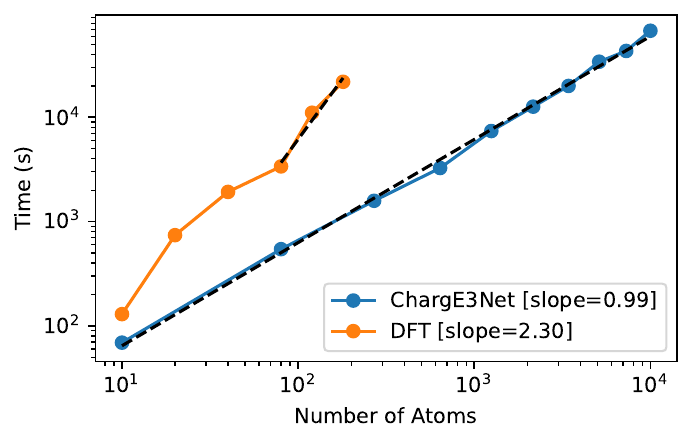}
    \caption{Runtime comparison of DFT and our \MODEL{} model with respect to number of atoms in the evaluated system. DFT is using a 48-core Intel Xeon CPU, while \MODEL{} uses a single NVIDIA V100 GPU.}
    \label{fig:runtime}
\end{figure}

\begin{figure}
    \centering
    \includegraphics[width=\linewidth]{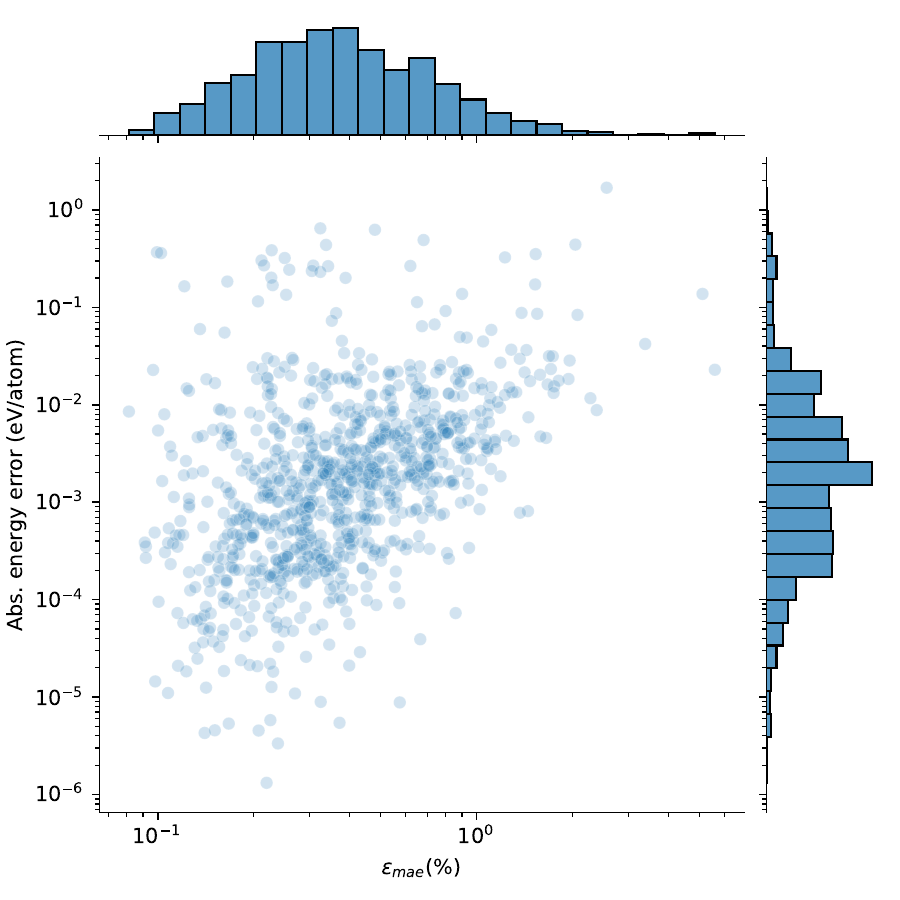}
    \caption{Per-atom energy absolute error between non-self-consistent calculations on \MODEL{}'s charge density predictions and self-consistent calculations plotted against $\epsilon_{mae}$ of the test set. The distributions of the per-atom energy error and $\epsilon_{mae}$ are on the right and top of the plot, respectively. Pearson $r\approx 0.44$ for $\log \epsilon_{mae}$ vs. log energy error.}
    \label{fig:per_atom_energy_error}
\end{figure}

\begin{figure}
    \centering
    \includegraphics[width=\linewidth]{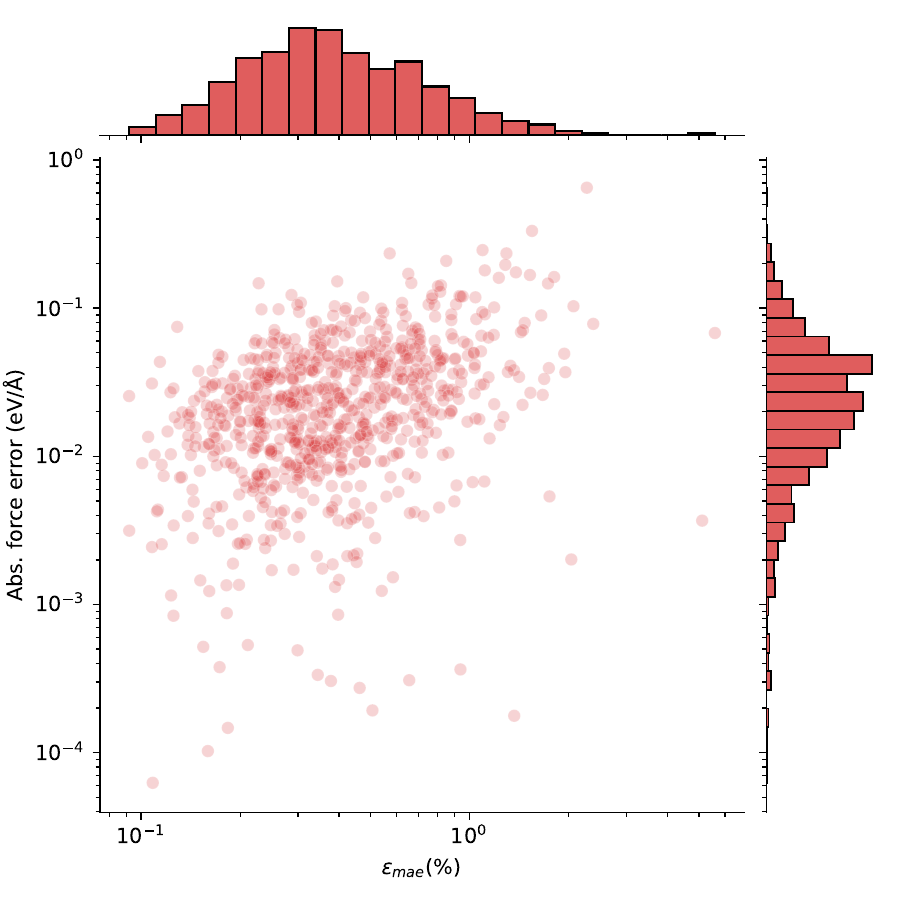}
    \caption{Average force absolute error between non-self-consistent calculations on \MODEL{}'s charge density predictions and self-consistent calculations plotted against $\epsilon_{mae}$ of the test set. The distributions of the force error and $\epsilon_{mae}$ are on the right and top of the plot, respectively. 15\% of structures achieved perfect agreement with the ground-truth self-consistent calculation, which are not shown in this plot. See \ref{fig:force_error_zero_hist} for the distribution of $\epsilon_{mae}$ of these structures. Pearson $r\approx 0.35$ for $\log \epsilon_{mae}$ vs. log force error.}
    \label{fig:force_error}
\end{figure}

\begin{figure}
    \centering
    \includegraphics[width=\linewidth]{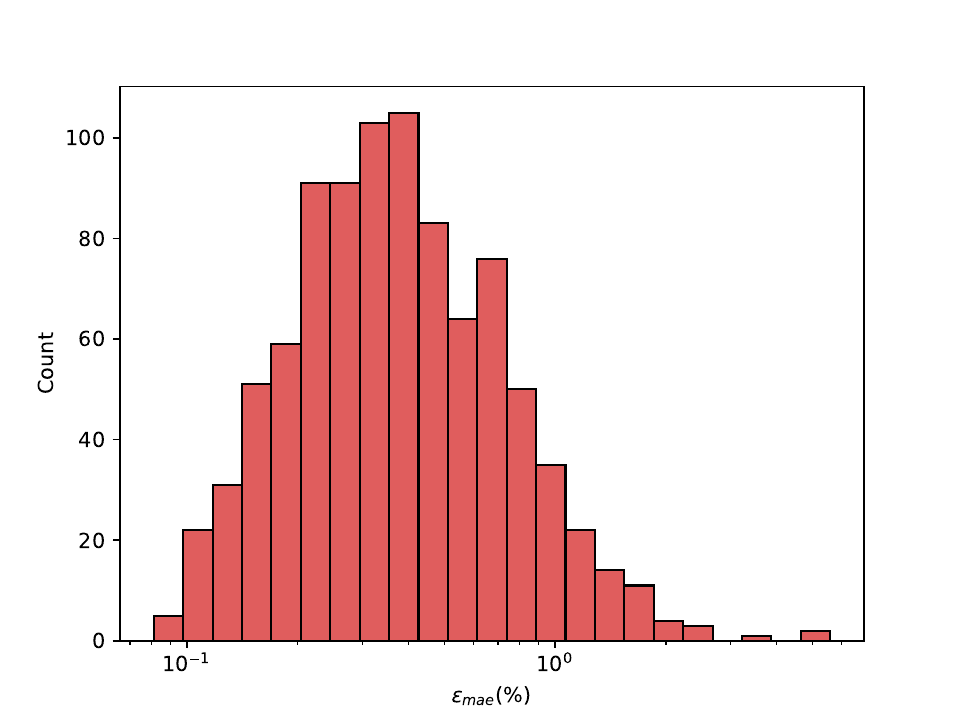}
    \caption{Distribution of $\epsilon_{mae}$ where each structure has a perfect agreement in forces between a non-self-consistent calculation initialized with \MODEL{}'s prediction and a self-consistent calculation.}
    \label{fig:force_error_zero_hist}
\end{figure}

\begin{figure}
    \centering
    \includegraphics[width=\linewidth]{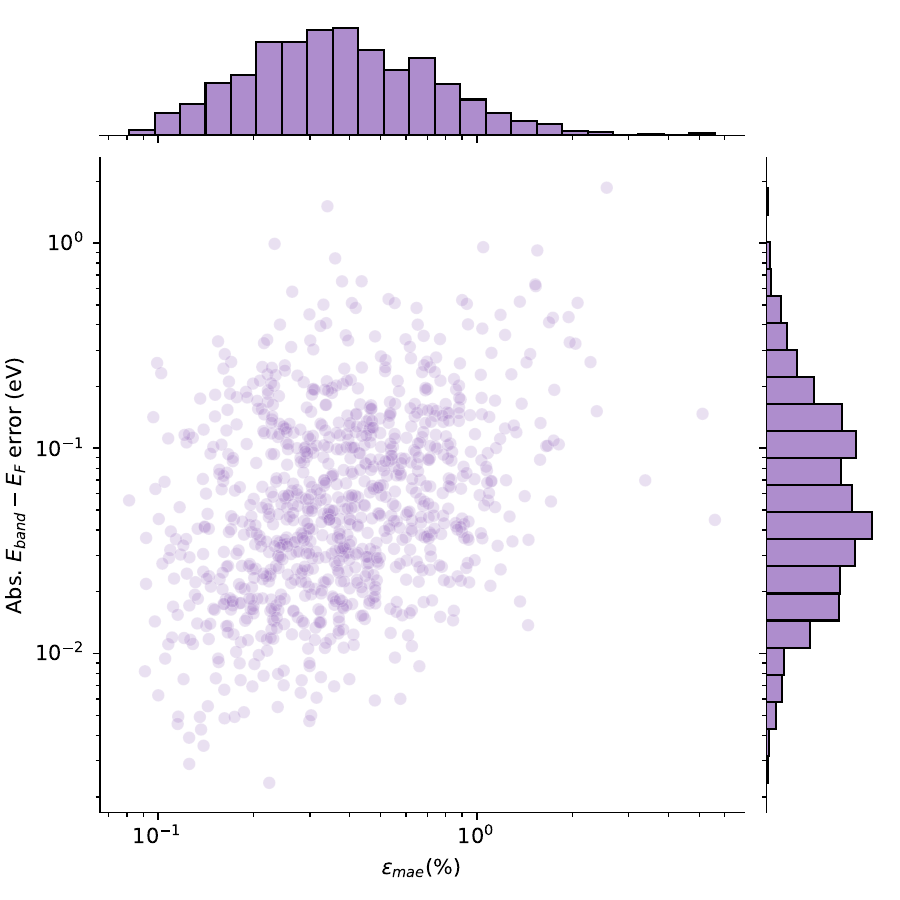}
    \caption{Band energy (eigenvalue) absolute error between non-self-consistent calculations on \MODEL{}'s charge density predictions and self-consistent calculations plotted against $\epsilon_{mae}$ of the test set. The distributions of the band energy error and $\epsilon_{mae}$ are on the right and top of the plot, respectively. Pearson $r\approx 0.35$ for $\log \epsilon_{mae}$ vs. log band energy error.}
    \label{fig:eigval_error}
\end{figure}

\begin{figure}
    \centering
    \includegraphics[width=\linewidth]{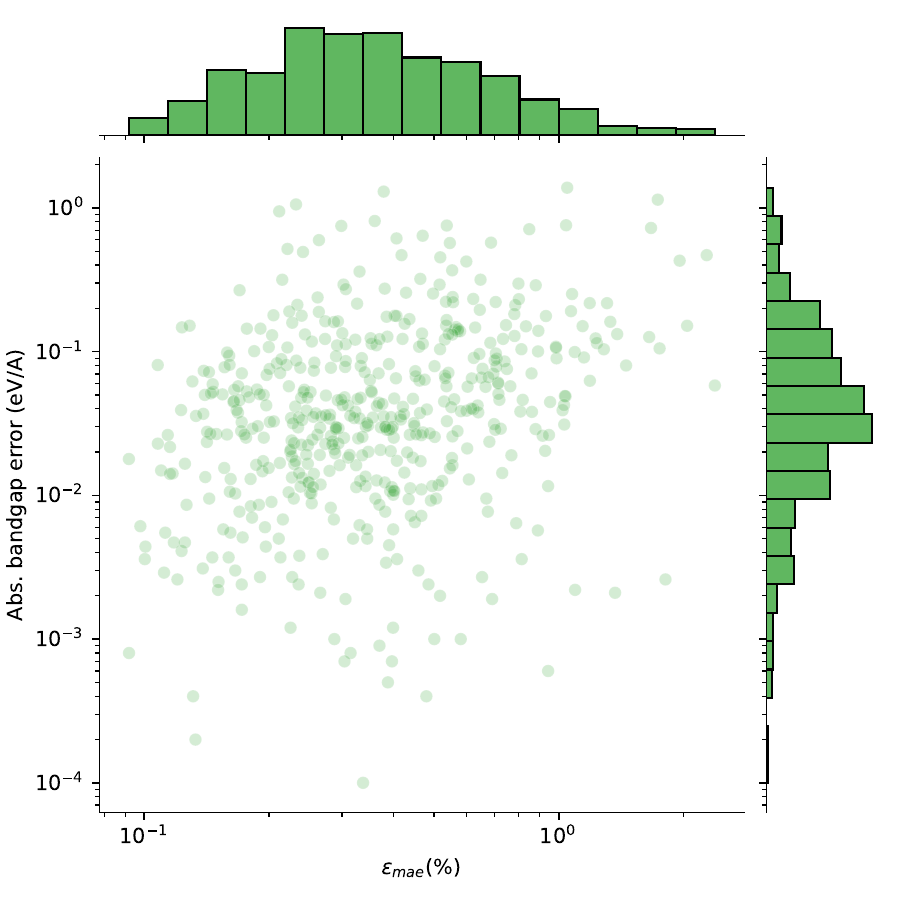}
    \caption{Band gap absolute error between non-self-consistent calculations on \MODEL{}'s charge density predictions and self-consistent calculations plotted against $\epsilon_{mae}$ of the test set. The distributions of the band gap error and $\epsilon_{mae}$ are on the right and top of the plot, respectively. About 43\% of structures achieve perfect band gap agreement with self-consistent calculations, which are not shown in this plot, and the distribution of their $\epsilon_{mae}$ is plotted in \ref{fig:gap_error_zero_hist}. Pearson $r\approx 0.29$ for $\log \epsilon_{mae}$ vs. log band gap error.}
    \label{fig:gap_error}
\end{figure}

\begin{figure}
    \centering
    \includegraphics[width=\linewidth]{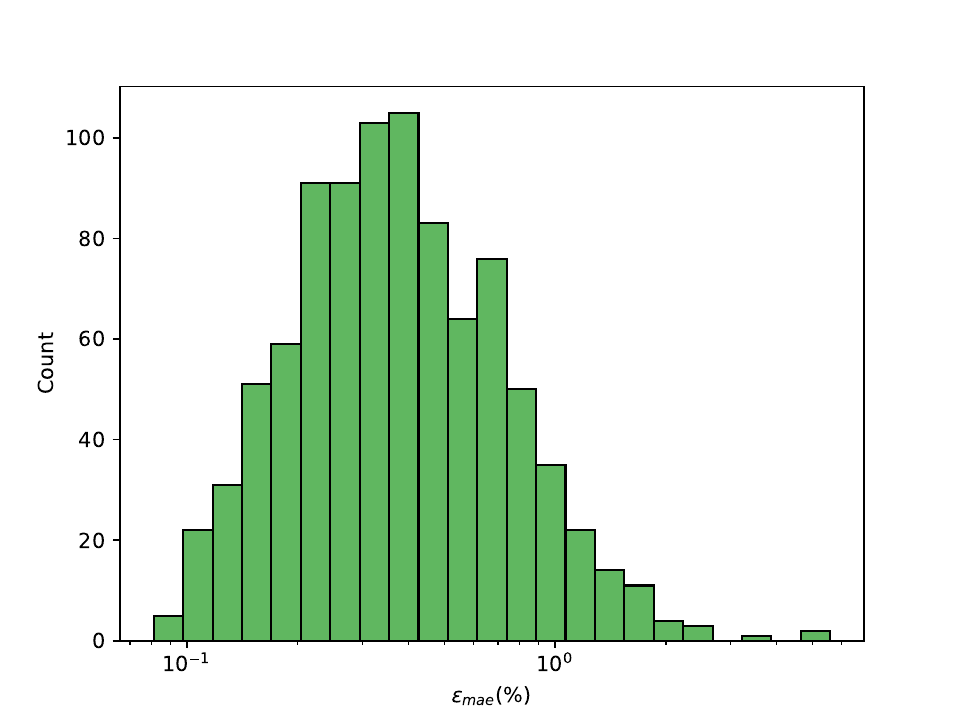}
    \caption{Distribution of $\epsilon_{mae}$ where each structure has a perfect band gap agreement between a non-self-consistent calculation initialized with \MODEL{}'s prediction and a self-consistent calculation.}
    \label{fig:gap_error_zero_hist}
\end{figure}

\begin{figure}
    \centering
    \includegraphics[width=\linewidth]{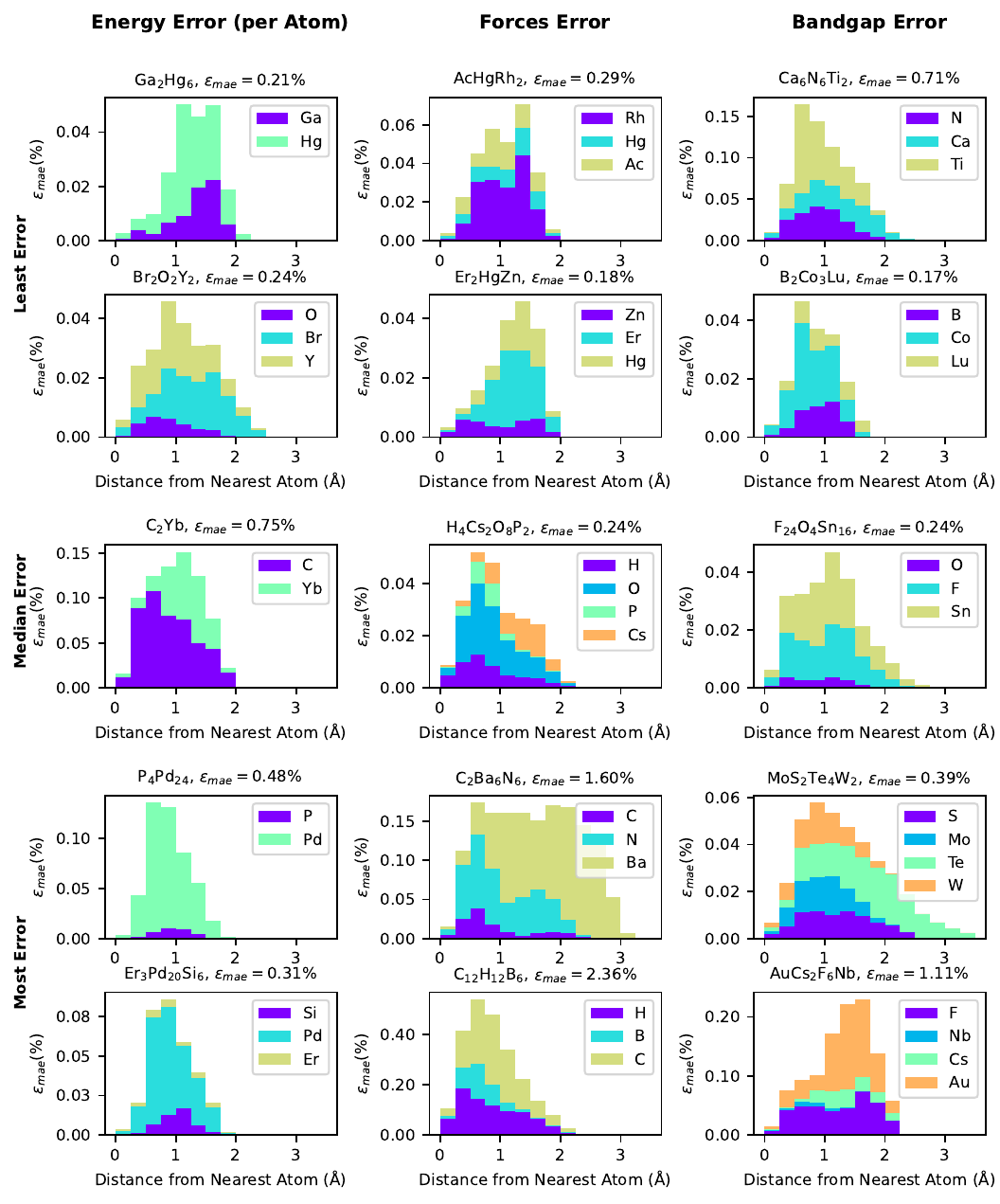}
    \caption{Stacked histograms showing $\epsilon_{mae}$, binned by range from nearest atom, for a selection of materials. $\epsilon_{mae}$ contributions from individual voxels on $G$ are summed in range bins and according to nearest atom species (stacked colors for each bin). Materials with the two least, median, and two most errors on each of three downstream NSC-from-\MODEL{} computed properties are shown in rows from top to bottom.}
    \label{fig:material_distance_error}
\end{figure}

\begin{figure}[!h]
  \includegraphics[width=0.98\linewidth]{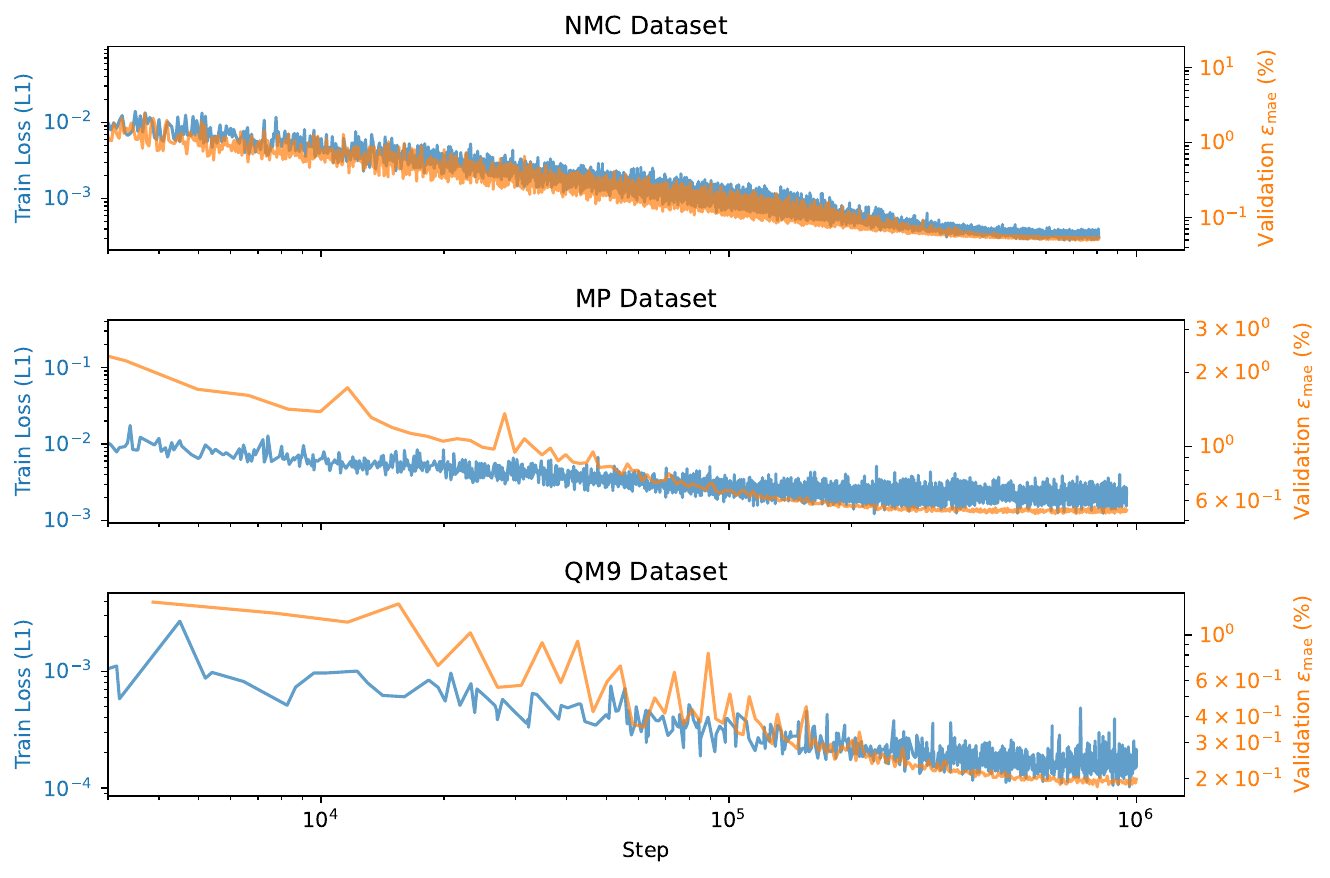}
  \caption{Log-log plot of training loss and validation $\epsilon_\text{mae}$ during training of the \MODEL{} for each of the studied datasets.}
  \label{fig:loss}
\end{figure}

\begin{figure}
    \centering
    \includegraphics[width=\linewidth]{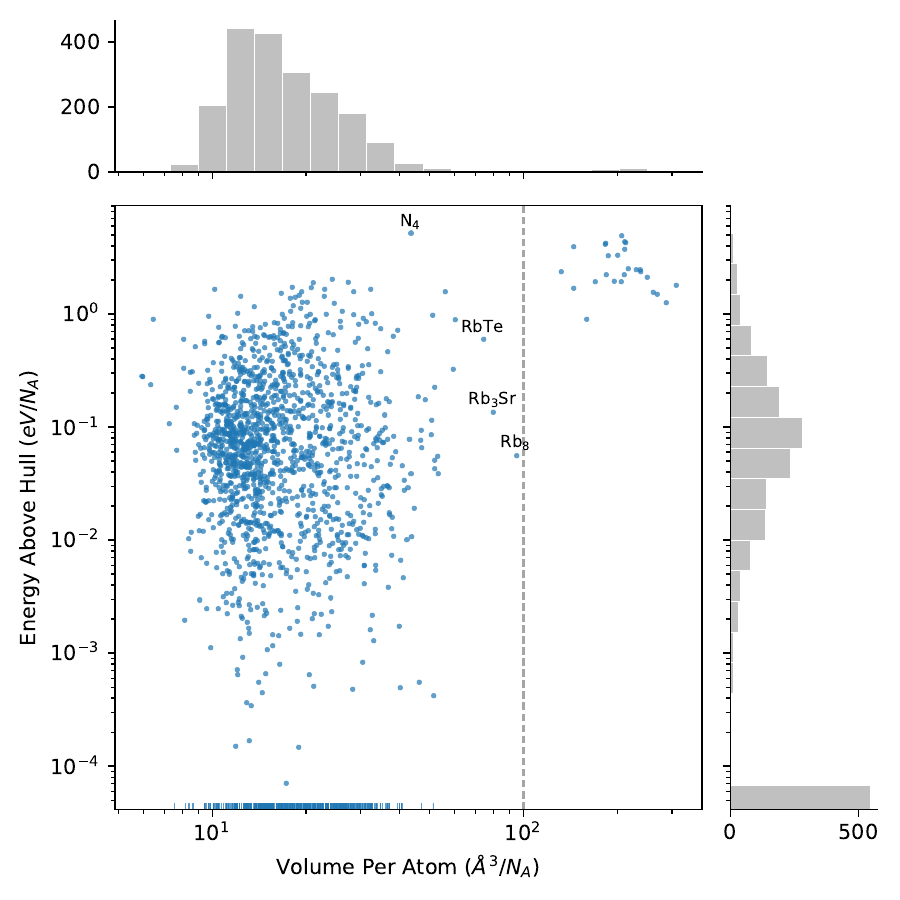}
    \caption{Scatterplot of test set Energy Above Hull vs Volume Per Atom ($n=2000)$. 26 materials in excess of 100 \AA$^3/N_A$ (dashed line) were removed from the test set for other calculations. Of those removed, 25 are in space group \textit{Immm} and one is in \textit{Cmmm}.  Thermodynamically stable materials ($0.0 eV/N_A$) are plotted along the x axis.}
    \label{fig:energy_volume}
\end{figure}

\end{document}